\documentclass[11pt]{article}

\newcommand{\calC}{{\cal C}}

\newcommand{\calR}{{\cal R}}
\newcommand{\calS}{{\cal S}}
\newcommand{\calU}{{\cal U}}

\newcommand{\real}{{\bf R}}
\def\eqref#1{(\ref{#1})} 
\newtheorem{theorem}{Theorem}[section]
\newtheorem{proposition}[theorem]{Proposition}
\newtheorem{lemma}[theorem]{Lemma}
\newtheorem{corollary}[theorem]{Corollary}
\newtheorem{remark}[theorem]{Remark}
\def\qed{\hbox{\hskip 6pt\vrule width6pt height7pt depth1pt
    \hskip1pt}\bigskip}  
\def\to{\rightarrow}

 \def\runinend{\enspace}
\def\ackname{Acknowledgement\runinend}%

\def\acknowledgements{\par\addvspace{17pt}\rmfamily
\def\ackname{Acknowledgements\runinend}%
\trivlist\if!\ackname!\item[]\else
\item[\hskip\labelsep
{\bf\ackname}]\fi}%

\begin{document} 
\bibliographystyle{plain} 

\hfill{} 
 
\thispagestyle{empty}

\begin{center}{\Large Asymptotic Behavior of Thermal Non-Equilibrium Steady States 
for a Driven Chain of Anharmonic Oscillators} 
\vspace{5mm}

Luc  Rey-Bellet\footnote{Department of Mathematics, Rutgers
University, 110 Frelinghuysen Road, Piscataway NJ 08854; Present address:
Department of Mathematics, University of Virginia, Kerchof Hall, 
Charlottesville VA 22903; 
lr7q@virginia.edu.}, Lawrence E. Thomas\footnote{Department of Mathematics,
University of Virginia, Kerchof Hall, Charlottesville VA 22903;
let@math.virginia.edu.}\\    
\vspace{5mm}

\begin{center}
{\small\it
$^1$ University of Rutgers, USA, \\
$^2$ University of Virginia, USA \\
}
\end{center}

\end{center}

\setcounter{page}{1} 
\begin{abstract} We consider a model of heat conduction introduced in
\cite{EPR1}, which consists of a finite nonlinear chain coupled to two
heat reservoirs at different temperatures. 
We study the low temperature asymptotic behavior of the invariant
measure. We show that, in this limit, the invariant measure is
characterized by a variational principle. We relate the heat flow to
the variational principle. The main technical ingredient is an
extension of Freidlin-Wentzell theory to a class of degenerate
diffusions. 
\end{abstract}

\section{Introduction} 
We consider a model of heat conduction introduced
in \cite{EPR1}. In this model a finite non-linear chain of $n$ $d$-dimensional 
oscillators is coupled to two Hamiltonian heat reservoirs initially at different 
temperatures $T_L$,$T_R$, and each of which is 
is described by a $d$-dimensional wave equation. 
A natural goal is to obtain a usable expression for the invariant (marginal) state of the 
chain analogous to the Boltzmann-Gibbs prescription $\mu= Z^{-1} \exp{(-H/T)}$ which 
one has in equilibrium statistical mechanics. What we show here is that the invariant state 
$\mu$ describing steady state energy flow through the chain is asymptotic to the expression 
$ \exp{(-W^{(\eta)}/T)}$ to leading order in the mean temperature $T$, $T\rightarrow 0$, 
where the action $W^{(\eta)}$, defined on phase space, is obtained from an explicit 
variational principle. 
The action $W^{(\eta)}$ depends on the temperatures only through the parameter 
$\eta=(T_L-T_R)(T_L+T_R)$. As one might anticipate, in the limit $\eta \rightarrow 0$,  
$W^{(\eta)}$ reduces to the chain Hamiltonian plus a residual term from the bath 
interaction, i.e., $\exp{(-W^{(\eta)}/T)}$ becomes
the Boltzmann-Gibbs expression. We remark that the variational principle for 
$W^{(\eta)}$ here certainly has analogues in more complicated arrays of oscillators, plates 
with multiple thermo-coupled baths, etc. The validity of this variational principle in more 
complex systems, as well as the physical phenomena to be deduced from $W^{(\eta)}$ are 
questions which remain to be explored.

Turning to the physical model at hand, we assume that the Hamiltonian $H(p,q)$ 
of the isolated chain is assumed to be of the form
\begin{eqnarray}
H(p,q)\,&=&\, \sum_{i=1}^n  \frac{p^2_i}{2} + \sum_{i=1}^{n} U^{(1)}(q_i)
+ \sum_{i=1}^{n-1} U^{(2)}(q_i-q_{i+1}) \,, \nonumber \\
\,&\equiv\,& \sum_{i=1}^n  \frac{p^2_i}{2} + V(q)\,, \label{hamil}
\end{eqnarray}
where $q_i$ and $p_i$ are the coordinate and momentum of the $i$-th particle, 
and where 
$U^{(1)}$ and $U^{(2)}$ are $\calC^\infty$ confining potentials, i.e. 
$\lim_{|q| \rightarrow \infty} V(q) = +\infty$. 

The coupling between the reservoirs and the chain is assumed to be of dipole 
approximation type and it occurs at the boundary only:
the first particle of the chain is coupled to one reservoir and the n-th 
particle to the other heat reservoir. At time $t=0$ each reservoir is assumed to be 
in thermal equilibrium, i.e., the initial conditions of the reservoirs are 
distributed 
according to (Gaussian) Gibbs measure with temperature $T_1=T_L$ and $T_n=T_R$ respectively.
Projecting the dynamics onto the phase space of the chain results in a set of 
integro-differential equations which differ from the Hamiltonian equations of motion by 
additional force terms in the equations for $p_1$ and $p_n$. Each of these terms consists 
of a deterministic integral part independent of temperature and a Gaussian random part with 
covariance proportional to the temperature. Due to the integral (memory) terms, the study of 
the long-time limit is a difficult mathematical problem (see \cite{JaPi} for the study of such 
systems in the case of a single reservoir).
But by a further  appropriate choice of couplings, 
the integral parts can be treated as auxiliary variables $r_1$ and $r_n$, the random parts 
become Markovian. 
Thus we obtain (see \cite{EPR1} for details) the following system  
of Markovian stochastic differential equations on the extended phase space 
$\real^{2dn + 2d}$: For $x=(p,q,r)$
\begin{eqnarray}
{\dot q_1} \,&=&\, p_1\,,  \nonumber\\
{\dot p_1} \,&=&\, -\nabla_{q_1}V(q) + r_1\,,  \nonumber\\
{\dot q_j} \,&=&\, p_j\,,  \nonumber\\
{\dot q_j} \,&=&\, -\nabla_{q_j}V(q)\,, \quad j=2,\dots,n-1\,, \nonumber\\
{\dot q_n} \,&=&\, p_n\,,  \nonumber\\
{\dot q_n} \,&=&\, -\nabla_{q_n}V(q) + r_n\,, \nonumber\\
d r_1 \,&=&\, -\gamma (r_1 - \lambda^2 q_1)dt  + (2\gamma\lambda^2
T_1)^{1/2} dw_1 \,, \nonumber\\
d r_n \,&=&\, -\gamma (r_n - \lambda^2 q_1)dt  + (2\gamma\lambda^2
T_n)^{1/2}  dw_n  \,, \label{eqmo002}
\end{eqnarray}
In Eq.~\eqref{eqmo002}, $w_1(t)$ and $w_n(t)$  are independent $d$-dimensional Wiener 
processes, and $\lambda^2$ and $\gamma$ are constants describing the couplings. 

It will be useful to introduce a 
generalized Hamiltonian $G(p,q,r)$ on the extended phase space, given by
\begin{equation}
G(p,q,r)\,=\, \sum_{i=1,n} \left( \frac{r^2_i}{2\lambda^2} - r_iq_i
\right) + H(p,q)\,, \label{ghamil} 
\end{equation}
where $H(p,q)$ is the Hamiltonian of the isolated systems of
oscillators given by \eqref{hamil}. We also introduce the 
parameters $\varepsilon$ (the mean temperature of the reservoirs) and $\eta$ (the 
relative temperature difference): 
\begin{equation}
\varepsilon\,=\, \frac{T_1 + T_n}{2}\,, \quad \eta\,=\, \frac{T_1-T_n}{T_1+T_n} \,.
\label{etaep}
\end{equation} 
Then Eq.~\eqref{eqmo002} takes  the form 
\begin{eqnarray}
{\dot q}\,&=&\, \nabla_p G  \,, \nonumber \\
{\dot p}\,&=&\, -\nabla_q G  \,, \nonumber \\
dr \,&=&\, -\gamma\lambda^2\nabla_r G  dt  + \varepsilon^{1/2} (2\gamma\lambda^2 D)^{1/2}
dw \,, \label{eqmo005}
\end{eqnarray}
where $p=(p_1,\dots,p_n)$,   $q=(q_1,\dots,q_n)$, $r=(r_1,r_n)$ and
where $D$ is the $2d \times 2d$ matrix given by
\begin{equation}
D\,=\,  \left( \begin{array}{cc} 1+\eta & 0 \\ 0
& 1-\eta \end{array}  \right)\,.
\end{equation}
The function $G$ is a Liapunov function, non-increasing in time, for the deterministic part of 
the flow \eqref{eqmo005}. 
If the system is in equilibrium, i.e, if $T_1=T_n=\varepsilon$ and $\eta=0$, 
it is not difficult to check that the generalized Gibbs measure 
\begin{equation}
\mu_{\varepsilon}=Z^{-1}\exp{(-G(p,q,r)/\varepsilon)} \,,
\label{eqm}
\end{equation} 
is an invariant measure for the Markov process solving Eq.~\eqref{eqmo005}.

If the temperature of the reservoirs are not identical, no explicit formula for 
the invariant measure $\mu_{T_1,T_n}$ can be given, in general. It is the goal of this paper 
to provide a variational principle for the leading asymptotic form for  $\mu_{T_1,T_n}$, at 
low temperature, 
$\varepsilon \rightarrow 0$.  To suggest what $\mu_{T_1,T_n}$ looks like, we observe
that a typical configuration of a reservoir has infinite 
energy, therefore the reservoir does not only acts as a sink of energy but true 
fluctuations can take place. The physical picture is as follows: the system spends most of the 
time very close to the critical set of $G$  (in fact close to a stable equilibrium) and very rarely 
(typically after an exponential time) an excursion far away from the equilibria occurs. 
This picture brings us into the framework of rare events, hence into the theory of 
large deviations and more specifically the Freidlin-Wentzell theory \cite{FW} of 
small random perturbations of dynamical systems.

In the following we employ notation which is essentially that of \cite{FW}. 
Let $\calC ([0,T])$  denote the Banach space of continuous functions (paths) with values 
in $\real^{2d(n+1)}$ equipped with the uniform topology. 
We introduce the  following functional $I^{(\eta)}_{x,T}$ on the set of paths $\calC([0,T])$: 
If $\phi(t)=(p(t),q(t),r(t))$ has one $L^2$-derivative with respect to time
and satisfies $\phi(0)=x$ we set
\begin{equation}
I_{x,T}^{(\eta)}(\phi) \,=\,  
\frac{1}{4\gamma\lambda^2} \int_0^T ({\dot r} +\gamma\lambda^2 \nabla_r G) D^{-1}
({\dot r} +  \gamma\lambda^2 \nabla_r G) dt \,,
\label{rf}
\end{equation}
if 
\begin{equation}
{\dot q}(t)= \nabla_pG(\phi(t))\,, \quad {\dot p}(t)=-\nabla_q G(\phi(t))\,,  
\label{constr}
\end{equation} 
and $I_{x,T}^{(\eta)}(\phi) = + \infty$ otherwise. Notice that $I_{x,T}^{(\eta)}(\phi)=0$ 
if and only if $\phi(t)$ is a solution of Eq.~\eqref{eqmo005} with the temperature $\varepsilon$ 
set equal to zero.
The functional $I^{(\eta)}_{x,T}$ is called a rate function and it describe, 
in the sense of large deviation, the probability of the path $\phi$: 
roughly speaking, as $\varepsilon \rightarrow 0$, 
the asymptotic probability of the path $\phi$ is given by 
\begin{equation}
\exp{ \left( - I_{x,T}^{(\eta)}(\phi)/\varepsilon \right)}\,.
\end{equation}
For $x,y \in \real^{2d(n+1)}$ we define $V^{(\eta)}(x,y)$ as
\begin{equation}
V^{(\eta)}(x,y) \,=\, \inf_{T>0} \inf_{\phi : \phi(T)=y}  I^{(\eta)}_{x,T}(\phi)\,, 
\label{popo}
\end{equation}
and for any sets $B$, $C \in \real^{2d(n+1)}$ we set
\begin{equation}
V^{(\eta)}(B,C) \,=\, \inf_{x\in B ; y\in C} V^{(\eta)}(x,y)\,.
\end{equation}
The function $V^{(\eta)}(x,y)$ represents, roughly speaking, the cost to bring the system from 
$x$ to $y$ (in an arbitrary amount of time). 
We introduce an equivalence relation on the phase space
$\real^{2d(n+1)}$: we say $x\sim y$ if $V^{(\eta)}(x,y)=V^{(\eta)}(y,x)=0$.
We divide the critical set $K=\{x\,;\, \nabla G(x)=0\}$ (about which the invariant measure  
concentrates) according to this equivalence relation: 
we have $K=\cup_{i} K_i$ with $x\sim y$  
if $x \in K_i, y \in K_i$ and $x\not\sim y $ if $x \in K_i, y \in K_j$, $i\not=j$.

Our first assumption is on the existence of an invariant measure, the structure of the 
set $K$ and the dynamics near temperature zero. Let $\rho >0$ be arbitrary 
and denote $B(\rho)$ the $\rho$-neighborhood of $K$ and let $\tau_\rho$ be the first time 
the Markov process $x(t)$ which solves \eqref{eqmo005} hits $B(\rho)$. 

\begin{itemize}
\item
{\bf K1} The process $x(t)$ has an invariant measure. 
The critical set $A$ of the generalized Hamiltonian $G$ can be decomposed 
into a finite number of inequivalent compact sets  $K_i$. 
Finally, for any $\varepsilon_0 >0$,  the expected hitting time 
$E_x(\tau_\rho)$ of the diffusion with initial condition $x$ is bounded uniformly for 
$0\le\varepsilon\le\varepsilon_0$.
\end{itemize}

\begin{remark}{\rm
The assumption {\bf K1} ensures that the dynamics is sufficiently 
confining in order to apply large deviations techniques to study the 
invariant measure.} 
\end{remark}

\begin{remark}{\rm
The assumptions used in \cite{EPR1,EH} to prove the existence of an 
invariant measure imply the assumption made on the structure of the critical set $A$. 
But it is not clear that they imply the assumptions made on the hitting time. We will 
merely assume the validity of condition {\bf K1} in this paper. Its validity can be 
established by constructing Liapunov-like functions for the model. Such methods allow 
as well to prove a fairly general theorem on the existence of invariant measures for 
Hamiltonian coupled to heat reservoirs (under more general conditions that in 
\cite{EPR1,EH}) and will be the subject of a separate publication \cite{RT}.
}
\end{remark}

Our second condition is identical to condition {\bf H2} of \cite{EPR1,EH}. 

\begin{itemize}
\item
{\bf K2} The 2-body potential $U^{(2)}(q)$ is strictly convex. 
\end{itemize}

\begin{remark}{\rm  
The condition {\bf K2} will be important to establish various 
regularity properties of $V^{(\eta)}(x,y)$. It will allow to imply several 
controllability properties of 
the control system associated with the stochastic differential equations \eqref{eqmo005}. }
\end{remark}

Following \cite{FW}, we consider graphs on the set $\{1,\dots,L\}$. 
A graph consisting of arrows $m\rightarrow n$, 
($m\in \{1,\dots,L\} \setminus \{i\}$, $m\in \{1,\dots,L\}$),   
is called a $\{i\}$-graph if 
\begin{enumerate}
\item Every point $j$, $j\not= i$ is the initial point of exactly one arrow. 
\item There are no closed cycles in the graph. 
\end{enumerate}
We denote $G\{i\}$ the set of $\{i\}$-graphs.
The weight of the set $K_i$ is defined by
\begin{equation}
W^{(\eta)}(K_i)\,=\, \min_{g\in G(\{i\})} \sum_{m\rightarrow n \in g} V^{(\eta)}(K_m,K_n)\,.
\label{ww}
\end{equation}

Our main result is the following: 
\begin{theorem} Under the conditions {\bf K1} and {\bf K2} the invariant measure 
$\mu_{T_1,T_n} = \mu_{\varepsilon,\eta}$ of the Markov process \eqref{eqmo005} has the 
following asymptotic behavior: For any open set $D$ with compact closure and 
sufficiently regular boundary 
\begin{equation}
\lim_{\varepsilon \rightarrow 0} \varepsilon \log \mu_{\varepsilon,\eta}(D) \,=\, 
\inf_{ x \in D} W^{(\eta)}(x)\,,
\label{asme}
\end{equation}
where
\begin{equation}
W^{(\eta)}(x)\,=\, \min_{i} \left( W^{(\eta)}(K_i) + V^{(\eta)}(K_i,x) \right) 
- \min_j W^{(\eta)}(K_j) \,.
\end{equation}
In particular, if $\eta=0$, then 
\begin{equation}
W^{(0)}(x)\,=\, G(x) - \min_x G(x)\,.
\label{eqq}
\end{equation}
The function $W^{(\eta)}(x)$ satisfies the bound, for $\eta \ge 0$, 
\begin{equation}
(1+\eta)^{-1} \left(G(x) - \min_x G(x)\right) \,\le\, W^{(\eta)}(x) \,\le\, 
(1-\eta)^{-1} \left(G(x) - \min_x G(x)\right)\,.
\label{neqq}
\end{equation}
\label{maintheorem}
and a similar bound for $\eta \le 0$. 
\end{theorem}

\begin{remark}{\rm 
Eqs. \eqref{eqq} and \eqref{neqq} imply that $\mu_{\varepsilon,\eta}$ reduces 
to the Boltzmann-Gibbs expression $\mu_{\varepsilon} \sim \exp{(-G/\varepsilon)}$ for 
$\eta \rightarrow 0$ in the low temperature limit. Of course, at $\eta=0$, they are 
actually equal at all temperatures $\varepsilon$. Moreover these equations imply that 
the relative probability $\mu_{\varepsilon,\eta}(x)/\mu_{\varepsilon,\eta}(y)$ is 
(asymptotically) bounded above and below by 
\begin{equation}
\exp{-\left[\frac{ G(x)}{\varepsilon(1 \pm \eta)} - 
\frac{G(y)}{\varepsilon(1 \mp \eta)} \right]}\,,
\end{equation}
so that no especially hot or cold spots 
develop for $\eta \not= 0$. 
}
\end{remark}

\begin{remark}{\rm 
The theorem draws heavily from the large deviations theory of  
Freidlin-Wentzell \cite{FW}. But the theory was developed for stochastic differential equations
with a non-degenerate (elliptic) generator, but for Eq.~\eqref{eqmo005} this is not the 
case since the random force acts only on $2d$  of the $2d(n+1)$ variables. A large 
part of this paper is devoted to simply extending Freidlin-Wentzell theory to a class
of Markov processes containing our model. Degenerate diffusions have been 
considered in \cite{DZ} but under too stringent conditions and also in \cite{BL} but 
with conditions quite different from ours. 
We also note that the use of Freidlin-Wentzell theory in non-equilibrium 
statistical mechanics has been advocated in particular by Graham (see \cite{Gr} and 
references therein).
In these applications to non-equilibrium statistical mechanics, as in \cite{Gr}, 
the models are mostly taken as  mesoscopic: the variables of the system describe some 
suitably coarse-grained quantities, which fluctuate slightly around their average values. 
In contrast to these models, ours is entirely microscopic and derived from first 
principles and the small-noise limit is seen as a low-temperature limit.
}
\end{remark}

Finally we  relate the large deviation functional to a kind of entropy 
production. As in \cite{EPR2} we define this entropy production $\Sigma$ by
\begin{equation}
\Sigma\,=\, -\frac{F_1}{T_1} -\frac{F_n}{T_n} \,,
\end{equation}
where
\begin{eqnarray}
F_1(p,q,r)\,&=&\,  p_1 (r_1 - \lambda^2 q_1) \,=\, \lambda^2 p_1 
\nabla_{r_1} G(p,q,r) \,, \nonumber \\
F_n(p,q,r)\,&=&\, p_n (r_n - \lambda^2 q_n) \,=\, \lambda^2 p_n 
\nabla_{r_n} G(p,q,r) \,,  
\end{eqnarray}
are the energy flows from the chain to the respective reservoirs. 
In \cite{EPR1} it is shown that $\mu_{T_1,T_n}(\Sigma) \ge 0$  and 
$\mu_{T_1,T_n}(\Sigma) = 0$ if and only if $T_1=T_n$. This implies that, 
in the stationary state, energy is flowing from the hotter reservoir bath to 
the colder one. With the parameters $\varepsilon$ and $\eta$ as defined in 
\eqref{etaep} we define $\Theta$  by
\begin{equation}
\frac{1}{\varepsilon} \Theta \,=\, \Sigma \, =\, \frac{1}{\varepsilon} 
\left( -\frac{F_1}{1+\eta} -\frac{F_n}{1-\eta} \right) \,.
\label{ep2}
\end{equation}
In order to show the relation between the rate function $I_{x,T}$ and the 
entropy production $\Sigma$, we introduce the time-reversal $J$, which is the 
involution on the phase space $\real^{2d(n+1)}$ given by $J(p,q,r)=(-p,q,r)$. 
The following shows that the value of the rate function of a path $\phi$ between 
$x$ and $y$ and is equal (up to a boundary term)
to the value of the rate function of the the time reversed path ${\tilde \phi}$ 
between $Jy$ and $Jx$ minus the entropy produced along this path. 
\begin{proposition}\label{ddd}
Let $\phi(t) \in \calC([0,T])$ with $\phi(0)=x$ and $\phi(T)=y$. 
Either $I^{(\eta)}_{x,T}(\phi) \,=\, + \infty$ or we have 
\begin{eqnarray}
I^{(0)}_{x,T}(\phi) \,&=&\, I^{(0)}_{Jy,T}({\widetilde \phi}) + 
G(y)- G(x)\,, \qquad {\rm~if~} \eta=0 \,, \nonumber \\
I^{(\eta)}_{x,T}(\phi) \,&=&\, I^{(\eta)}_{Jy,T}({\widetilde \phi}) + 
R(y)- R(x)  -
\int_0^T  \Theta(\phi(s)) ds\,, {\rm~~~if~} \eta\not= 0\,,  \label{dtbl}
\end{eqnarray}
where $\Theta$ is defined in Eq.~\eqref{ep2} and 
$R(x)=(1+\eta)^{-1}(\lambda^{-1}r_1-\lambda q_1)^2 + 
(1-\eta)^{-1}(\lambda^{-1}r_n-\lambda q_n)^2$. 
\end{proposition}
The identities given in Proposition \ref{ddd} are an asymptotic version of  
identities which appear in various forms in the literature. These identities are the 
basic ingredient needed for the proof of the Gallavotti-Cohen fluctuation theorem  
\cite{ECM,GC} for stochastic dynamics \cite{Kur,LeSp,Ma} and appear as well 
in the Jarsynski non-equilibrium work relation \cite{Ja}.

The paper is organized as follows: In Section \ref{large} we recall the large 
deviation principle for the paths of Markovian stochastic differential equation and 
using methods from control theory we prove the required regularities properties of the 
function $V^{(\eta)}(x,y)$ defined in Eq.~\eqref{popo}. Section \ref{asym} is devoted to an 
extension of Freidlin-Wentzell results to a certain class of diffusions with 
hypoelliptic generators (Theorem \ref{single}): 
we give a set of conditions under which the asymptotic behavior
of the invariant measure is proved. The result of Section \ref{large} implies that 
our model, under Assumptions {\bf K1} and {\bf K2}, satisfies the conditions of 
Theorem \ref{single}. In Section \ref{balance} 
we prove the equality \eqref{eqq} and the bound \eqref{neqq} which depend on the 
particular properties of our model.

\section{Large deviations and Control Theory}\label{large}

In this section we first recall a certain numbers of concepts and theorems which will
be central in our analysis:   The large deviation principle for the sample path of 
diffusions introduced by Schilder for the Brownian motion \cite{Sc} and 
generalized to arbitrary diffusion by \cite{FW,Az,Va} (see also \cite{DZ}),  
and the relationship between diffusion processes and control theory,
exemplified by the Support Theorem of Stroock and Varadhan \cite{SV}.
With these tools we then prove several properties of the dynamics for our model. 
We prove that ``at zero temperature'' the (deterministic) dynamics given by  
is dissipative: 
the $\omega$-limit set is the set of the critical point of $G(p,q,r)$. We also prove 
several properties of the control system associated to Eq.~\eqref{eqmo005}: 
a local control property 
around the critical points of $G(p,q,r)$ and roughly speaking 
a global ``smoothness'' property of the weight of the paths
between $x$ and $y$, when $x$ and $y$ vary. The central hypothesis in this analysis is 
condition {\bf K2}: this condition implies the hypoellipticity, \cite{Ho}, 
of the generator of the  Markov semigroup associated to Eq.~\eqref{eqmo005}, but 
it implies in fact a kind of global hypoellipticity which will be used here to prove 
the aforementioned properties of the dynamics.

\subsection{Sample Paths Large Deviation and Control Theory}\label{ldpct}
Let us consider the stochastic differential equation
\begin{equation}\label{stodif}
dx(t)\,=\, Y(x) dt + \varepsilon^{1/2} \sigma(x) dw(t)\,,
\end{equation}
where $x\in X= \real^n$, $Y(x)$ is a $\calC^\infty$ vector field,
$w(t)$ is an m-dimensional Wiener 
process and  $\sigma(x)$ is a $\calC^\infty$ map 
from $\real^m$ to $\real^n$. 
Let $\calC ([0,T])$ denote the Banach space of continuous functions
with values in $\real^n$ equipped with the uniform topology.  
Let $L^2([0,T])$ denote the set of square integrable
functions with values in $\real^m$ and $H_1([0,T])$ denote
the space of absolutely continuous functions with values in $\real^m$ 
with square integrable derivatives.
Let $x_\varepsilon(t)$ denote the solution of \eqref{stodif}
with initial condition $x_\varepsilon(0)=x$. We assume that $Y(x)$ and
$\sigma(x)$ are 
such that, for arbitrary $T$, the paths of the
diffusion process $x_\varepsilon(t)$  belong to $\calC ([0,T])$. 
We let $P^\varepsilon_{x}$ denote 
the probability measure on $\calC ([0,T])$ induced by
$x_\varepsilon(t)$, $0\le t \le T$ and denote $E^\varepsilon_{x}$ the corresponding 
expectation.  

We introduce the rate function $I_{x,T}(f)$ on $\calC ([0,T])$ given by
\begin{equation}
I_{x,T}(f)\,=\, \inf_{\{g\in H_1 : f(t)=x+\int_0^T Y(f(s)) ds +
\int_0^T\sigma(f(s)) {\dot g}(s) ds \}} \frac{1}{2} \int_0^T |{\dot
g}(t)|^2 dt \,, \label{ratef}
\end{equation}
where, by definition, the infimum over an empty set is taken as
$+\infty$. The rate function has a particularly convenient form for us
since it accommodates degenerate situations where ${\rm rank}\, \sigma <
n$.

In \cite{DZ}, Corollary 5.6.15 (see also \cite{Az}) the following
large deviation principle
for the sample paths of the solution of \eqref{stodif} is proven. It
gives a version of the large deviation principle which is uniform in
the initial condition of the diffusion.

\begin{theorem} \label{LDP}
Let $x^{\varepsilon}(t)$ denote the solution of
Eq.~\eqref{stodif} with initial condition $x$. 
Then, for any $x \in \real^n$ and for any $T < \infty$, the rate
function $I_{x,T}(f)$ is a lower semicontinuous function on $\calC([0,T])$ 
with compact level sets (i.e. $\{ f\,;\, I_{x,T}(f) \le \alpha
\}$ is compact for any $\alpha \in \real$). 
Furthermore the family of measures $P^\varepsilon_{x}$
satisfy the large deviation principle on $\calC ([0,T])$ with rate
function $I_{x,T}(f)$: 
\begin{enumerate}
\item For any compact $K\subset X$ and any closed $F\subset \calC
([0,T])$,
\begin{equation}
\limsup_{\varepsilon \rightarrow 0} \log \sup_{x\in K} P_x(x_{\varepsilon}
\in F ) \,\le\, - \inf_{x\in K} \inf_{\phi \in F} I_{x,T}(\phi). 
\end{equation}
\item For any compact $K\subset X$ and any open $G\subset \calC
([0,T])$,
\begin{equation}
\liminf_{\varepsilon \rightarrow 0} \log \inf_{x\in K} P_x(x_{\varepsilon}
\in G ) \,\ge\, - \sup_{x\in K} \inf_{\phi \in G} I_{x,T}(\phi). 
\end{equation}
\end{enumerate}
\end{theorem}

Recall that for our model given by Eq.~\eqref{eqmo005}, the rate function 
takes the form given in Eqs. \eqref{rf} and \eqref{constr}.  
We introduce further the cost function $V_T(x,y)$ given by
\begin{equation}\label{fcost}
V_T(x,y)\,=\,\inf_{\phi \in \calC ([0,T])\,:\, \phi(T)=y} I_{x,T}(\phi)\,.
\end{equation}
Heuristically $V_T(x,y)$ describes the cost of forcing the system to
be at $y$ at time $T$  starting from $x$ at time $0$.
The function $V(x,y)$ defined in the introduction, Eq. \eqref{popo} is equal to
\begin{equation}\label{icost}
V(x,y)\,=\,\inf_{T>0} V_T(x,y)\,,
\end{equation}
and describes the minimal cost of forcing the system from $x$ to
$y$ in an arbitrary amount of time. 
 
The form of the rate function suggest a connection between large deviations 
and control theory. In Eq.~\eqref{ratef}, the infimum is taken over functions 
$g\in H_1([0,T])$ which are more regular than a path of the 
Wiener process. If we 
do the corresponding substitution in Eq.~\eqref{stodif}, we obtain an ordinary 
differential equation 
\begin{eqnarray}
{\dot x}(t)\,&=&\, Y(x(t)) + \sigma(x(t)) u(t) \label{cteq} \,,
\end{eqnarray}
where we have set $u(t)= \varepsilon^{1/2}{\dot g}(t) \in L^2([0,T])$.  
The map $u$ is called a control and the equation \eqref{cteq} a control system. 
We fix an arbitrary time $T >0$.
We denote by $\varphi_x^u:[0,T ]\to \real^n$ the solution  
of the differential equations \eqref{cteq}  with control $u$ and initial
condition $x$.
The correspondence between the stochastic system Eq.~\eqref{stodif} and
the deterministic system Eq.~\eqref{cteq} is exemplified by the Support Theorem 
of Stroock and Varadhan \cite{SV}.
The support of the diffusion process $x(t)$ with initial condition $x$
on $[0,T]$, 
is, by definition, the smallest closed subset $\calS_x$ of $\calC
([0,T])$ such that 
\begin{equation}\label{bigP}
P_x [ x(t) \in \calS_x ]\,=\, 1 ~.
\end{equation}
The Support Theorem asserts that the support of the diffusion is equal to the 
set of solutions of Eq.~\eqref{cteq} as the control $u$ is varied: 
\begin{equation}
\calS_x\,=\, { \overline { \{ \varphi_x ^u \, : \, u \in L^2([0,T]) ~\} } }~,
\label{allxt}
\end{equation}
for all $x\in\real^k$. 
The control system \eqref{cteq} is said to be {\it strongly completely controllable}, if 
for any  $T>0$, and any pair of points $x,y$, there exist a control $u$ such that 
$\varphi_x^u(0)=x$ 
and $\varphi_x^u(T)=x$. In \cite{EPR2} it is shown that, under condition {\bf K2}, 
the control system associated with the equation \eqref{eqmo005} is  strongly completely 
controllable. This is an ergodic property and this implies, \cite{EPR2}, uniqueness 
of the  invariant measure (provided it exists). 
In terms of the cost function $V_T(x,y)$ defined 
in \eqref{fcost}, strong complete controllability simply means that  $V_T(x,y) < \infty$, 
for any $T>0$ and any $x,y$. 
The large deviation principle, Theorem \ref{LDP}, gives
more quantitative 
information on the actual weight of paths between $x$ and $y$ in time
$T$,  in particular that the weight is $\sim
\exp(-\frac{1}{\varepsilon} V_T(x,y))$. As we will see below, these weights
will determine completely the leading (exponential) behavior of the
invariant measure for $x_{\varepsilon}(t)$, $\varepsilon \downarrow 0$.

\subsection{Dissipative properties of the dynamics}\label{dissipative}

We first investigate the $\omega$-limit set of the dynamics ``at temperature
zero'', i.e,  when both temperatures $T_1$, $T_n$ are set equal to
zero in the equations of motion. In this case the dynamics is
deterministic and, as the following result shows, dissipative.   

\begin{lemma}
Assume condition {\bf K2}. Consider the system of  differential equations 
given by 
\begin{eqnarray}
{\dot q}_i &=&  \nabla_{p_i} G \qquad i=1,\cdots,n  \,,
\nonumber \\ 
{\dot p}_i &=& - \nabla_{q_i} G \qquad  i=1,\cdots,n
\,, \label{zerom}  \\ 
{\dot r}_i &=& - \gamma \lambda^2 \nabla_{r_i} G
\qquad i=1,n \,. \nonumber  
\end{eqnarray} 
Then the $\omega$-limit set of the flow given by Eq.(\ref{zerom}) is the
set of critical points of the generalized Hamiltonian 
$G(p,q,r) = \sum_{j=1,n}(\lambda^{-2}r_{j}^{2}/2 -r_{j}q_{j})
+H(p,q)$, i.e., 
\begin{equation}
A\,=\, \left\{ x\in {\bf R}^{2d(n+1)}\,:\, \nabla G(x)=0 \right\}\,.
\end{equation}
\label{omegalemma}
\end{lemma}

\noindent {\em Proof:} As noted in the introduction $G(x)$ is a Liapunov function for the
flow given by (\ref{zerom}). A simple computation shows that
\begin{eqnarray}
\frac{d}{dt} G(x(t))\,&=& \, -\gamma \lambda^2 \sum_{i=1,n}
(\lambda^{-2}r_i(t) - q_i(t))^2 
\nonumber \\
             \,&=&\, - \gamma\lambda^2 \sum_{i=1,n} |\nabla_{r_i} G(x(t))|^2
\, \leq \, 0 \,. \label{liap} 
\end{eqnarray}
Therefore it is enough to show that the flow does not get ``stuck'' at
some point of the hyper-surfaces $(\lambda^{-2}r_i(t) - q_i(t))^2=0, i=1,n$
which does not belong to the set $A$.  

Let us assume the contrary, i.e., that, for some trajectory and some
times $T_1 < T_2$ we have 
\begin{equation}
G(x(t))\,=\, G(x(T_1))\, {\rm ~for~}  t \in [T_1,T_2]\,.
\label{const}
\end{equation}
We show that this implies that $x(t)\in A$, for $t \in
[T_1,T_2]$. From Eqs.~\eqref{liap} and \eqref{const} we 
have, for  $t \in [T_1,T_2]$, the identity
\begin{equation}
\lambda^{-2}r_1(t)-q_1(t)\,=\,  \nabla_{r_1} G(x(t))\,=\, 0 \,.
\end{equation}
Further, using Eq.(\ref{zerom}), we obtain
\begin{equation}
0 \,=\, \frac{ d}{ dt}
(\lambda^{-2}r_1(t)-q_1(t)) \,=\, - \gamma (\lambda^{-2}r_1(t)-q_1(t))
- p_1(t)\,=\, -  p_1(t)\,.
\end{equation}
Thus we get
\begin{equation}
p_1(t)\,=\, \nabla_{p_1}G(x(t))\,=\, 0\,.
\end{equation}
Since $p_1(t)$ is a constant, 
\begin{equation}
0\,=\, -{\dot p_1}(t) \,=\, \nabla_{q_1}G(x(t)) \,=\,
\nabla_{q_1}V(q(t)) - r_1(t) \,,
\label{induc}
\end{equation} 
and therefore
\begin{equation}
 \nabla_{q_1}G(x(t))\,=\, 0 \,.
\end{equation}
Using that $\lambda^{-2}r_1(t)-q_1(t)=0$ we can rewrite Eq.~\eqref{induc} as
follows:
\begin{equation}
0\,=\, \nabla_{q_1}U^{(1)}(q_1(t)) - \lambda^{2}q_1(t) + \nabla_{q_1}
U^{(2)}(q_1(t)-q_2(t))\,. 
\label{induc2}
\end{equation}  
By the convexity condition on $U^{(2)}$, {\bf K2}, $\nabla U^{(2)}: \real^d
\rightarrow \real^d$ is a diffeomorphism with an inverse which we
denote $W$. We can therefore solve Eq.(\ref{induc2}) in 
terms of $q_2$ and we obtain
\begin{equation}
q_2(t)\,=\, q_1(t) - W \left(\lambda^2 q_1(t) -
\nabla_{q_1}U^{(1)}(q_1(t))\right) \,\equiv
\, F(q_1(t))\,.  
\end{equation}
From this we conclude that
\begin{equation}
p_2(t)\,=\,{\dot q_2}(t)\,=\, \nabla_{q_1}F(q_1(t)) {\dot
q}_1(t)\,=\,\nabla_{q_1}F(q_1(t))p_1(t)\,=\,0 \,,  
\end{equation}
and thus
\begin{equation}
p_2(t)\,=\,  \nabla_{p_2} G(x(t))\,=\, 0 \,. 
\end{equation}
Proceeding by induction along the chain it is easy to see that if
$G(x(t))$ is 
constant on the interval $[T_1,T_2]$, then one has
\begin{equation}
 \nabla G(x(t))\,=\,0\,,
\end{equation}
and therefore $x(t) \in A$. This concludes the proof of Lemma
\ref{omegalemma}. \qed

\subsection{Continuity properties of $V^{(\eta)}_T(x,y)$}\label{uppercont}

In the analysis of the asymptotic behavior of the invariant measure it will be
important to establish certain continuity properties of the cost function
$V^{(\eta)}_T(x,y)$. We prove first a global property: we show that for any
time $T$, $V^{(\eta)}_T(x,y)$ as a map from $X\times X \rightarrow \real$ is
everywhere finite and upper semicontinuous. Furthermore we need a
local property of $V^{(\eta)}_T(x,y)$ near the $\omega$-limit set of the
zero-temperature dynamics (see Lemma \ref{omegalemma}). We prove that
if $x$ and $y$ are sufficiently close to this $\omega$-limit set then
$V^{(\eta)}_T(x,y)$ is small. Both results are obtained using control theory and
hypoellipticity. 

\begin{proposition} \label{continuity}
Assume condition {\bf K2}. Then the functions $V^{(\eta)}_T$, for all $T>0$ and  
$V^{(\eta)}$ are  upper semicontinuous  maps : $X\times X \rightarrow \real$. 
\end{proposition}

\noindent
{\em Proof:} By definition $V^{(\eta)}_T(y,z)$ is given by
\begin{equation}\label{cost}
V^{(\eta)}_T(y,z)\,=\, \inf \frac{1}{2} \int_0^T \sum_{j=1,n} |u_j(t)|^2 dt \,,
\end{equation}
where the infimum in \eqref{cost} is taken over all $u=(u_1,u_n) \in
L^2([0,T]) $ such that
\begin{eqnarray}
{\dot q}\,&=&\, \nabla_p G \,,\nonumber \\
{\dot p}\,&=&\, -\nabla_q G \,, \nonumber \\
{\dot r}\,&=&\, -\gamma\lambda^2 \nabla_r G + (2\gamma\lambda^2 D)^{1/2}  
u \,,\label{eqmo17}
\end{eqnarray}
with boundary conditions
\begin{equation}
(p(0),q(0),r(0))=y\,, \qquad  (p(T),q(T),r(T))=z \,. 
\end{equation}
In other words, the
infimum in \eqref{cost} is taken over all controls $u$ which steer
$y$ to $z$. 
For notational simplicity we let $r_1=q_0$ and
$r_n=q_{n+1}$. Furthermore we set 
$Q=(q_0,q_1,\dots q_n,q_{n+1}) \in \real^{d(n+2)}$ and 
$P=({\dot q_1}, \dots, {\dot q_n}) \in \real^{dn}$.   
The equations \eqref{eqmo17} take the form
\begin{eqnarray}
{\dot q}_0 \,&=&\, -\gamma \lambda^2 \nabla_{q_0}G(q_0,q_1) +
(2\gamma\lambda^2 (1+\eta))^{1/2} u_1 \,,
\nonumber \\
{\ddot q}_l \,&=&\, -\nabla_{q_j}G(q_{l-1}, q_{l}, q_{l+1}) \quad
l=1,\dots, n \,, \nonumber \\
{\dot q}_{n+1} \,&=&\, -\gamma \lambda^2
\nabla_{q_{n+1}}G(q_n,q_{n+1})  
+ (2\gamma\lambda^2 (1-\eta))^{1/2} u_n \,, \label{eqmo33} 
\end{eqnarray}
with boundary conditions 
\begin{equation}
\left( P(0), Q(0) \right) \,=\, y\,, \quad 
\left( P(T), Q(T) \right) \,=\,z\,. \label{bc01}
\end{equation}
By condition {\bf K2}, $\nabla_{q} U^{(2)}(q)$ is a diffeomorphism. As
a consequence the identity 
\begin{equation}
{\ddot q}_l \,=\, -\nabla_{q_l}G(q_{l-1}, q_{l}, q_{l+1})
\end{equation}
can be solved for either $q_{l-1}$ or $q_{l+1}$: there are
smooth functions $G_l$ and $H_l$ such that
\begin{eqnarray}
q_{l-1}\,&=&\, G_l (q_l,{\ddot q}_l, q_{l+1}) \,, \label{inv1} \\
q_{l+1}\,&=&\, H_l (q_{l-1}, q_l,{\ddot q}_l) \,.  \label{inv2}
\end{eqnarray}
Using this we rewrite now the equations in the following form:
We assume for simplicity $n$ is an even number and we set $j=n/2$. (If
$n$ is odd, take $j=(n+1)/2$ and up to minor modifications the
argument goes as in the even case).

We rewrite Eq.~\eqref{eqmo33} as follows: For the first $j+1$ equations we use
Eq.~\eqref{inv1} and find
\begin{eqnarray}
u_1\,&=&\, \frac{1}{(2\gamma\lambda^2 (1+\eta))^{1/2}} ({\dot q}_0
+\gamma \lambda^2 \nabla_{q_0}G(q_0,q_1)) \, \equiv \, G_0(q_0,{\dot
q}_0, q_1) \,, \nonumber \\
q_l\,&=&\, G_{l+1}(q_{l+1},{\ddot q}_{l+1}, q_{l+2}) \quad l=0,1,\dots,
j-1 \,. \label{eqmo36}
\end{eqnarray}
For the remaining $n+1-j=j+1$ equations we use Eq.~\eqref{inv2} and obtain
the equivalent equations
\begin{eqnarray}
u_n\,&=&\,\frac{1}{(2\gamma\lambda^2 (1-\eta))^{1/2}} ({\dot q}_{n+1}
+\gamma \lambda^2 \nabla_{q_{n+1}}G(q_n,q_{n+1})) \nonumber  \\
\, &\equiv& \,  H_{n+1}(q_n, q_{n+1}, {\dot q}_{n+1}) \,, \nonumber \\
q_l&=& H_{l-1}(q_{l-2}, q_{l-1}, {\ddot q}_{l-1}) \quad
l=j+2,\dots, n+1 \,. \label{eqmo37}
\end{eqnarray}
Obviously both sets of equations \eqref{eqmo36} and \eqref{eqmo37} 
can be solved iteratively to express $u_1, q_0,\dots, q_{j-1}$ and
$q_{j+2}, \dots, q_{n+1}, u_n$ as functions of  only $q_j$, $q_{j+1}$
and a certain number of their derivatives.  
We note $q^{[\alpha]}=(q,q^{(1)}, \dots, q^{(\alpha)})$ where
$q^{(k)} = d^k q/dt^k $. From Eq.~\eqref{eqmo36} we obtain, for some
smooth functions $I_0,\dots I_{j}$, the set of equations 
\begin{eqnarray}
u_1\,&=&\, I_0 \left( q_j^{[2j+1]}, q_{j+1}^{[2j-1]} \right) \,, \label{u0} \\
q_k\,&=&\, I_{k+1}\left(q_j^{[2(j-k)]}, q_{j+1}^{[2(j-1-k)]} \right)\,, 
\quad k=0,1,\dots,j-1\,.   \label{qlj}
\end{eqnarray}
Similarly from Eq.~\eqref{eqmo37}, we find  smooth functions
$J_0,\dots J_{j}$ such that	
\begin{eqnarray}
q_k\,&=&\, J_{n+2-k} \left( q_j^{[2(k-j-2)]}, q_{j+1}^{[2(k-j-1)]} \right)\, 
\quad k=j+2 \dots, n+1 \,. \label{qgj} \\
u_n\,&=&\, J_{0} \left( q_j^{[2j-1]}, q_{j+1}^{[2j+1]} \right) \,. \label{un} 
\end{eqnarray}
So far we have simply rewritten the differential equations of motion
in an implicit form. From this we can draw the following conclusions. 
If $\left( P(t), Q(t) \right)$ is  
a solution of
Eq.~\eqref{eqmo33} with given control $u=(u_1,u_n)$, then, using Eqs.~\eqref{u0} and 
\eqref{un} $u$ can be
written as follows: there is a smooth function $B$ such that
\begin{equation} 
u(t)\,=\,  B \left( q_j^{[2j+1]}(t),  q_{j+1}^{[2j+1]}(t)\right) 
\label{control} \,,
\end{equation}
i.e., $u$ can be expressed as a function of the functions $q_j$, $q_{j+1}$ 
and their first $2j+1$ derivatives. Furthermore if  
$ \left( P(t), Q(t) \right)$ is a solution of
Eq.~\eqref{eqmo33}, from Eqs.~\eqref{qlj} and \eqref{qgj}, 
we can express $\left( P, Q \right)$ 
as a function of  $q_j$, $q_{j+1}$ and their first $2j$ derivatives. 
In particular this defines a map 
$N : \real^{2d(n+1)} \rightarrow \real^{2d(n+1)}$, 
where $ \left( P(t), Q(t) \right) = N \left( q_j^{[2j+1]}, q_{j+1}^{[2j+1]} 
\right)$ 
is given by
\begin{eqnarray}
q_k\,&=&\, I_{k+1} \left(q_j^{[2(j-k)]}, q_{j+1}^{[2(j-1-k)]} \right)\,, 
\quad k=0,1,\dots,j-1\,,   \nonumber \\
{\dot q}_k\,&=&\, {\dot I}_{k+1}\left(q_j^{[2(j-k)+1]}, q_{j+1}^{[2(j-1-k)+1]}\right)\,, 
 \quad k=1,\dots,j-1\,,   \nonumber \\
q_k\,&=&\,q_k\,, \quad k=j,j+1\,, \nonumber \\
{\dot q}_k\,&=&\,{\dot q}_k\,, \quad k=j,j+1\,, \nonumber \\
q_k\,&=&\,  J_{n+2-k}\left(q_j^{[2(k-j-2)]}, q_{j+1}^{[2(k-j-1)]}\right)\,, 
\quad k=j+2 \dots, n+1 \,, \nonumber \\
{\dot q}_k\,&=&\, {\dot J}_{n+2-k}\left(q_j^{[2(k-j-2)+1]}, 
q_{j+1}^{[2(k-j-1)+1]} \right)\,, \quad k=j+2 \dots, n \,. \label{nn}
\end{eqnarray}
We now show that $N$ is a homeomorphism, by constructing explicitly  
its inverse. 
We use the equations of motion \eqref{eqmo33} to derive equations for 
$q_j^{[2j]}$ and $ q_{j+1}^{[2j]}$. Differentiating repeatedly the equations 
with respect to time one inductively finds functions smooth functions 
$K_0, \dots K_{2j}$ and $L_0, \dots L_{2j}$ such that
\begin{eqnarray}
q_j^{(k)}\,&=&\, 
   K_k\left( q_0^{[0]}, q_1^{[1]}, \dots, q_{j-1}^{[1]}, q_{j}^{[k-2]}, 
         q_{j+1}^{[k-2]}\right)\,, \quad k=0,\dots, 2j\,, \label{qjk} \\
q_{j+1}^{(k)}\,&=&\, 
   L_k \left( q_{j}^{[k-2]}, q_{j+1}^{[k-2]}, q_{j+2}^{[1]}, \dots, 
         q_n^{[1]}, q_{n+1}^{[0]} \right)\,, \quad k=0,1, \dots, 2j \,.
\label{qj+1k}
\end{eqnarray}
Eqs.~\eqref{qjk} and \eqref{qj+1k} define
inductively a smooth map $M$  from  $\real^{2d(n+1)}$ to  $\real^{2d(n+1)}$ given by
\begin{equation}
\left(q_j^{[2j]}, q_{j+1}^{[2j]}\right)\,=\, M(P,Q)\,. \label{hom2}  
\end{equation}
We have shown that if 
$\left( P(t), Q(t) \right)$ is a solution of Eq.~\eqref{eqmo33}, then 
\begin{equation}
\left(q_j^{[2j]}(t), q_{j+1}^{[2j]}(t)\right) = M \left( P(t), Q(t) \right)\,.
\end{equation}
Since the solution of \eqref{eqmo33} is unique this shows that $M$ is 
the inverse of  the map $N$ given by Eq.~\eqref{nn} and thus $N$ is a 
homeomorphism (in fact a diffeomorphism).  

We have proven the following: The system of equations \eqref{eqmo33}
with boundary data \eqref{bc01} is equivalent to equation \eqref{control} 
with the boundary data    
\begin{eqnarray}
\left( q_j^{[2j]}(0), q_{j+1}^{[2j]}(0)\right) \,=\, M(y)\,, \quad 
\left( q_j^{[2j]}(T), q_{j+1}^{[2j]}(T)\right) \,=\, M(z)\,. \label{bc02} 
\end{eqnarray}
From this the assertion of the theorem follows easily: First we see
that $V^{(\eta)}_T(y,z) < \infty$, for all $T>0$ and for all $y,z$. 
Indeed choose any sufficiently smooth curves $q_j(t)$ and
$q_{j+1}(t)$ which satisfies the boundary conditions
\eqref{bc02} and consider the $u$ given by Eq.~\eqref{control}. 
Then the function $\left(P(t),Q(t)\right)= 
M\left( q_j^{[2j]}(t), q_{j+1}^{[2j]}(t)\right)$
is a solution of Eq.~\eqref{eqmo33}  with boundary data \eqref{bc01} and 
with a control $u(t)$ given by \eqref{control} which steers $y$ to $z$.

In order to prove the upper semicontinuity of $V^{(\eta)}_T(y,z)$, let us choose some
$\epsilon >0$. By definition of $V^{(\eta)}_T$ there is a  control $u$ which
steers $y$ to $z$ along a path $\phi=\phi^u$ such that 
\begin{equation}
I_{y,T}(\phi^u)\,\le \, V^{(\eta)}_T(y,z) + \epsilon/2 \,,
\end{equation} 
and 
\begin{equation} 
u(t)\,=\, B\left(  q_j^{[2j+1]}(t), q_{j+1}^{[2j+1]}(t)\right) \,,
\end{equation}
Using the smoothness of $B$, we choose curves ${\tilde q}_j$ and
${\tilde q}_{j+1}$ such that 
\begin{equation}
\sup_{t\in [0,T]} | q^{[2j+1]}_j  -{\tilde q}^{[2j+1]}_j | + |
q^{[2j+1]}_{j+1}  -{\tilde q}^{[2j+1]}_{j+1} | \, \le \, \delta \,,
\end{equation} 
and $\delta$ is so small that
\begin{eqnarray}
{\tilde u}(t) \,=\,B( {\tilde q}_j^{[2j+1]}, {\tilde q}_{j+1}^{[2j+1]})\,,
\end{eqnarray}
satisfies
\begin{equation} 
\sup_{t\in [0,T]} |u(t)-{\tilde u}(t)| \, \le \sqrt{\frac{\epsilon}{T}}\,.
\end{equation}
We note ${\tilde y} = M( {\tilde q}_j^{[2j]}(0), {\tilde q}_{j+1}^{[2j]}(0))$ 
and  ${\tilde z} = N( {\tilde q}_j^{[2j]}(T), {\tilde q}_{j+1}^{[2j]}(T))$ and 
${\phi}^u$ the path along which the control $u$ steers
the system, one obtains
\begin{equation}
|I_{y,T}(\phi^u) - I_{{\tilde y},T}(\phi^{\tilde u})|  \, \le \, \frac{\epsilon}{2} \,.
\end{equation}
By the continuity of the map $N$, we
can choose $\delta'$ so small that if $|y-{\tilde y}| + |z-{\tilde z}|
\le \delta'$, then 
\begin{eqnarray}
&&| q^{[2j]}_j(0)  -{\tilde q}^{[2j]}_j(0) | + |
q^{[2j]}_{j+1}(0)  -{\tilde q}^{[2j]}_{j+1}(0) | 
\nonumber \\
&&+ | q^{[2j]}_j(T)
-{\tilde q}^{[2j]}_j(T) | + | 
q^{[2j]}_{j+1}(T)  -{\tilde q}^{[2j]}_{j+1}(T) | \le \delta \,.\nonumber
\end{eqnarray}
Therefore for all such ${\tilde y},{\tilde z}$ we have
\begin{equation}
V^{(\eta)}_T({\tilde y},{\tilde z}) \,\le \, I_{{\tilde y},T}( \phi^{\tilde u}) \, \le
\, V^{(\eta)}_T(y,z) + \epsilon\,. 
\end{equation} 
This shows the upper semicontinuity of $V^{(\eta)}_T(y,z)$ and the upper 
semicontinuity of $V^{(\eta)}(y,z)$ follows easily from this. This
concludes the proof of Lemma \ref{continuity}. \qed

An immediate consequence of this Lemma is a bound  on the cost
function around critical points of the generalized Hamiltonian $G$. 

\begin{corollary} \label{small}
For any $x\in A =\{ y\,:\, \nabla G(y)=0\}$ and any 
$h>0$ there is $\delta >0$ such that, if $|y-x| + |z-x| \le \delta$,
then one has
\begin{equation}
V^{(\eta)}(y,z)\,\le \, h \,.
\end{equation}
\end{corollary}

\noindent
{\em Proof:} If $x\in A$, $x$ is a stationary point of the equation
\begin{eqnarray}
{\dot q}\,&=&\, \nabla_p G \,,\nonumber \\
{\dot p}\,&=&\, -\nabla_q G \,, \nonumber \\
{\dot r}\,&=&\, -\gamma\lambda^2 \nabla_r G \,. \label{eqmo53}
\end{eqnarray}
As a consequence the control $u\equiv 0$ steers $0$ to $0$ and hence
$V^{(\eta)}(x,x)=0$. The upper semicontinuity of $V^{(\eta)}(y,z)$ immediately implies the
statement of the corollary. \qed

\begin{remark}{\rm
This corollary slightly falls short of what is needed to obtain the
asymptotics of the invariant measure. More detailed information about
the geometry of the control paths around the stationary points is
needed and will be proved in the next subsection. 
}
\end{remark}

\subsection{Geometry of the paths around the stationary points}\label{localcont}

Let us consider a control system of the form
\begin{equation}\label{co1}
{\dot x} \,=\, Y(x) + \sum_{i=1}^m X_i(x) u_i
\end{equation}
where $x\in \real^n$, $Y(x), X_i(x)$ are smooth vector fields. 
We assume that $Y(x), X_i(x)$ are such that
Eq.~\eqref{co1} has a unique solution for all time $t>0$.
We want to investigate properties of the set which can be reached from
a given point by allowing only controls with bounded size. The class
of controls $u$ we consider is given by 
\begin{equation}
\calU_M\,=\, \left\{ u {\rm~piecewise~smooth},{\rm~ with~}
|u_i(t)| \le M ~\,,~ 1\le i \le m \right\}~.
\end{equation}
We denote $Y_{\le \tau}^M(x)$ the set of points which can be reached
from $x$ in time less than $\tau$ with a control $u \in \calU_M$. 
We say that the control system is {\em small-time locally
controllable} (STLC) at $x$ if $Y_{\le \tau}^M(x)$ contains a
neighborhood of $x$ 
for every $\tau > 0$. 

The following result is standard in control theory, see e.g. \cite{Su} or
\cite{LM} for a proof.  

\begin{proposition} \label{local}
Consider the control system Eq.~\eqref{co1} with $u\in
\calU_M$. Let $x_0$ be an equilibrium point of $Y(x)$, i.e.,
$Y(x_0)=0$. If the linear span of the brackets
\begin{equation}
{\rm ad}^k (Y)(X_i)(x) \quad i=1,\dots, m\,, \quad k=0,1,2,\dots \,,
\end{equation}
has rank $n$ at $x_0$ then  Eq.~\eqref{co1} is STLC at $x_0$.
\end{proposition}

\noindent
{\em Proof:} One proves Lemma \ref{local} by linearizing around $X_0$ 
and using e.g. the implicit function theorem, see e.g. \cite{LM}, Chapter 6, 
Theorem 1.  \qed

As a consequence of Lemma \ref{local} and results obtained in \cite{EPR1} one 
gets

\begin{lemma}\label{gcont} Consider the control system given by Eqs. \eqref{eqmo17}
%\begin{eqnarray}
% {\dot q}\,&=&\, \nabla_p G\,, \nonumber \\
% {\dot p}\,&=&\, - \nabla_q G \,, \nonumber \\
% {\dot r}\,&=&\, - \gamma \lambda^2\nabla_r G + (2\gamma \lambda^2 D)^{1/2} u \,,\label{c1}
% \end{eqnarray}
with $u \in \calU_M$. Let $x_0$ be a critical point of $G(x)$. If condition 
{\bf K2} is satisfied, then the system \eqref{eqmo17} is STLC at $x_0$.
\end{lemma}

\noindent
{\em Proof:} The property of small time local controllability is expressed 
as a condition that certain brackets generate the whole tangent space at some 
point $x_0$.  This property is obviously related  
to the hypoellipticity of the 
generator of the Markov process \eqref{eqmo005} associated to the control system 
\eqref{eqmo17}. The generator of the Markov process which solves $dy(t)=
Y(x) + \sum_i X_i(x) dw_i(t)$, where $w_i(t)$ is a $1$-dimensional process is 
given on sufficiently smooth functions by the differential operator $L = (1/2)
\sum_i (\nabla\cdot X_i)(X_i\cdot\nabla) + Y\cdot\nabla$. If $Y(x)$ and 
$X_i(x)$ are $\calC^\infty$, then $L$ is hypoelliptic if the Lie algebra 
generated by $Y(X)$ and $X_i(x)$ generates the tangent space at each point $x$ \cite{Ho}.
For the system of equations \eqref{eqmo005}, it is proved in \cite{EPR1}, that 
if condition {\bf K2} is satisfied, then the brackets 
\begin{equation}
{\rm ad}^k (Y)(X_i)(x) \quad i=1,\dots, m\,, \quad k=0,1,2,\dots 
\end{equation}
generates the tangent space at each point $x$, in particular at 
every critical point $x_0$, and therefore by Lemma \ref{local}, the control 
system Eq.~\eqref{eqmo17} is STLC at $x_0$. \qed

With these results we can derive the basic fact on the geometry of the
control paths around equilibrium points of $G(x)$.

\begin{proposition} \label{cont}
Consider the control system given by \eqref{eqmo17}.
Let $x_0$ be a critical point of $G(x)$ and $B(\rho)$ the ball of radius 
$\rho$ centered at $X_0$.
Then for any $h>0$, there are $\rho' >0$ and $\rho> 0$ with
$\rho <\rho'/3$ such that the following hold: For any $x,y \in
B(\rho)$, there is $T>0$ and  $u \in \calU_M$ with
\begin{equation}
\phi^u(0)=x\, \qquad\phi^u(T)=y\,,
\end{equation}
\begin{equation}
\phi^u(t) \in B(2\rho'/3)\,,  \qquad  t\in [0,T]\,,
\end{equation}
and 
\begin{equation}
I_{x,T}(\phi^u) \, \le \, h \,.
\end{equation}
\end{proposition}

\noindent
{\em Proof:} Together with the control system \eqref{eqmo17}, we consider
the time-reversed system
\begin{eqnarray}
{\dot {\tilde q}}\,&=&\, - \nabla_p G\,, \nonumber \\
{\dot {\tilde p}}\,&=&\,  \nabla_q G \,, \nonumber \\
{\dot {\tilde r}}\,&=&\,  \gamma \lambda^2\nabla_r G + (2\gamma\lambda^2 D)^{1/2} u
\label{c2}\,. 
\end{eqnarray}
Lemma \ref{gcont} implies the STLC of the control system
\eqref{eqmo17}. Furthermore from Lemma \ref{local} it is easy to see the
control system \eqref{c2} is STLC if and only if the control system
\eqref{eqmo17} is. We note $\phi^u$ (${\tilde \phi}^u$) the solution of
Eq.~\eqref{eqmo17} (Eq.~\eqref{c2}) and $Y^M_T(x)$  (${\tilde Y}^M_T(x)$)
the set of reachable points for the control system \eqref{eqmo17}
(\eqref{c2}). 
Using the convexity of the set of values the control can assume and
the continuous dependence of $\phi^u$ on $u$, it is easy to see
(\cite{Su}, Prop. 2.3.1) that $Y^M_T(x)$ is a compact set. 

We choose now $M$ and $T$ such that $M^2T \le h$. Since $Y^M_T(x)$
and ${\tilde Y}^M_T(x)$ are compact, there is $\rho' >0$ such that 
\begin{equation}
Y^M_T(x), {\tilde Y}^M_T(x) \subset B(2\rho'/3)\,.
\end{equation}
Furthermore we may choose $\rho'$ arbitrarily small by choosing 
$M$ and/or $T$ sufficiently small. 
By Lemma \ref{gcont}, both systems \eqref{eqmo17} and \eqref{c2} are
STLC and thus there is $\rho >0$ with $\rho < \rho'/3$ and 
\begin{equation}
B(\rho) \subset Y^M_T(x), {\tilde Y}^M_T(x) \,.
\end{equation}
Therefore there are controls $u_1, u_2 \in \calU_M$ such that
\begin{eqnarray}
\phi^{u_1}(0)\,=\,x_0 \,, &\quad& \phi^{u_1}(T)\,=\,y  \,,   \label{co5} \\
{\widetilde \phi}^{u_2}(0)\,=\,x_0 \,, &\quad& {\widetilde
\phi}^{u_2}(T)\,=\,x  \,. \label{co6}  
\end{eqnarray}
By reversing the time, the trajectory ${\widetilde \phi}^{u_2}(t)$ yields a
trajectory $\phi^{u_2}(t)$ with $\phi^{u_2}(0)=x$ and $\phi^{u_2}(T)=x_0$. 
Concatenating the trajectories $\phi^{u_2}(t)$ and $\phi^{u_1}(t)$ yields a
path $\phi$ from $x$ to $y$ which does not leave the ball
$B(2\rho'/3)$ and 
for which we have the estimate
\begin{equation}
I_{x,2T}(\phi)\,=\,  \frac{1}{2} \int_0^{2T} dt  |u(t)|^2  \,\le\, M^2T
\,\le\, h \,, 
\end{equation}
and this concludes the proof of Corollary \ref{cont}. \qed

\section{Asymptotics of the invariant measure}\label{asym}

In this section we prove an extension of Freidlin-Wentzell theory \cite{FW} 
for a certain class of diffusion processes with hypoelliptic generators 
concerning the invariant measure. 
Such extensions, for the problem of the exit from a 
domain, exist, see \cite{BL}, where a strong hypoellipticity 
condition is assumed which is not satisfied in our model 
and see also \cite{DZ}, where their assumption of small-time local 
controllability on the boundary of the domain is too strong for our purposes. 
Once the control theory estimates have been established, 
our proof follows rather closely the proof of 
Freidlin-Wentzell \cite{FW} and the presentation of it given in \cite{DZ} 
with a number of technical modifications.   

We consider a stochastic differential equation of the form
\begin{equation}
dx_{\varepsilon} \,=\, Y(x_\varepsilon) + \varepsilon^{1/2} \sigma(x_\varepsilon) dw\,,
\label{eqmo00}
\end{equation}
where $x\in X=\real^n$,  $Y(x)$ is a $\calC^\infty$ vector field, 
$\sigma(x)$ a  $\calC^\infty$ map from $\real^m$ to $\real^n$ and $w(t)$ 
a standard $m$-dimensional Wiener process. We view the stochastic process given 
by Eq.~\eqref{eqmo00} as a small perturbation of the dynamical system
\begin{equation}
{\dot x}\,=\, Y(x) \,.
\label{eqmo01}
\end{equation}
We denote $I_{x,T}(\cdot)$ the large deviation functional associated to 
Eq.~\eqref{eqmo00} (see Eq.~\eqref{ratef}) and denote $V_T(x,y)$ and $V(x,y)$ 
the cost functions given by \eqref{fcost} and \eqref{icost}. 
As in \cite{FW} we introduce an equivalence relation $\sim$ on $X$ 
defined as follows: $x \sim y$ if $V(x,y)=V(y,x)=0$. 

Our assumptions on the diffusion process $x_\varepsilon(t)$ are the following
\begin{itemize}
\item{{\bf L0}} The process $x_\varepsilon(t)$ has an invariant measure 
$\mu_\varepsilon$. 
\item{{\bf L1}} There is a finite number of compact sets $K_1, K_2, \dots K_L$ such that
\begin{enumerate} 
\item For any two points $x,y$ belonging to the same $K_i$ we have $x\sim y$.
\item If $x\in K_i$ , $y\in K_j$, with $i\not= j$, then $x \not\sim y$. 
\item Every $\omega$-limit set of the dynamical system \eqref{eqmo01} is contained 
in  $K_i$. 
\end{enumerate}
We let $B(\rho)$ denote the $\rho$ neighborhood of $\cup_i K_i$ and $\tau_\rho$ the first 
time the diffusion $x_\varepsilon(t)$ hits the set $B(\rho)$.
We assume that for any $\varepsilon_0 >0$  the expected hitting time 
$E_x(\tau_\rho)$ of the diffusion with initial condition $x$ is bounded uniformly for 
$0\le\varepsilon\le\varepsilon_0$.
\item{{\bf L2}} The diffusion process $x_\varepsilon(t)$ has an 
hypoelliptic generator.  Moreover, for any $x\in K_i$ the control system associated 
to Eq.~\eqref{eqmo00} is small-time locally controllable. 
\item{{\bf L3}} The diffusion process is strongly completely controllable, i.e., for all 
$T>0$, $V_T(x,y) < \infty$ and, moreover, $V_T(x,y)$ is upper semicontinuous  as a 
map from $X\times X $ to $\real$. 
\end{itemize}

\begin{remark}{\rm 
For the model we consider, condition which ensures that {\bf L0} holds 
are given in \cite{EPR1,EH}. For condition {\bf L1} we assume that the set of 
critical points of $G(p,q,r)$ is a compact set and item 
(iii) follows from Lemma \ref{omegalemma}. The bound on the expected hitting time will be 
proved in a separate publication \cite{RT}. 
For condition {\bf L2}, the hypoellipticity of the generator 
and the small-time local controllability follows from {\bf K2}, 
see Lemmas \ref{local}, \ref{gcont} and \ref{cont}. 
Condition {\bf L3} is a consequence of condition {\bf K2}, see Proposition 
\ref{continuity}.  
}
\end{remark}

\begin{remark}{\rm 
Condition {\bf L2} is a {\it local} property of the dynamics and as such sufficient 
conditions can be given in terms of adequate  Lie algebra.  
A simple sufficient condition 
for small-time local controllability was quoted and used in Section \ref{localcont}. 
More general sufficient conditions have been proved, see \cite{Su} and references 
therein. condition {\bf L3} is a {\it global} condition on the dynamics and we are 
not aware of any general condition which would imply {\bf L3} (except of course 
ellipticity of the generator). 
}
\end{remark}

To describe  the asymptotic behavior of the invariant measure $\mu_\varepsilon$ we 
will need the following quantities. We let
\begin{eqnarray} 
V(K_i,K_j) \,&=&\, \inf_{y\in K_i, z\in K_j} V(y,z)\,,  \quad  i,j=1,\dots, L \,, 
\label{vij} \\ 
V(K_i,z) \,&=&\, \inf_{y\in K_i} V(y,z)\,,  \quad  i=1,\dots,  L \,.
\label{viz}  
\end{eqnarray} 
We set 
\begin{equation}
W(K_i)\,=\, \min_{g \in G\{i\}} \sum_{(m\rightarrow n) \in g} V(K_m,K_n) \,, \label{wi}
\end{equation}  
where the set of $\{i\}$-graphs $G\{i\}$ is defined in the paragraph above Theorem 
\ref{maintheorem}. 
The asymptotics of the invariant measure is 
given by the function $W(x)$ given by
\begin{equation}
W(x)\,=\, \min_{i} \left( W(K_i) + V(K_i,x) \right) - \min_j W(K_j) \,.
\label{wx}
\end{equation}

We call a domain $D\subset X$ regular, if the boundary of $D$,
$\partial D$ is a piecewise smooth manifold. Our main result is the following:

\begin{theorem}\label{single}
Assume conditions {\bf L0-L3}. Let $D$ be a regular domain with compact closure 
such that ${\rm dist}(D,\cup_i K_i) >0$. 
Then the (unique) invariant measure $\mu_\varepsilon$  of the process $x_\varepsilon(t)$ 
satisfies 
\begin{equation} \label{mm}
\lim_{\varepsilon \rightarrow 0} \varepsilon \ln \mu_\varepsilon(D) \,=\, 
- \inf_{z\in D} W(z) \,.
\end{equation}
In particular if there is a single critical set $K$ one has
\begin{equation} \label{mms}
\lim_{\varepsilon \rightarrow 0} \varepsilon \ln \mu_\varepsilon(D) \,=\, 
- \inf_{z\in D} V(K,z) \,.
\end{equation}
\end{theorem}

We first recall some general results on hypoelliptic diffusions obtained in 
\cite{Kl}, in particular a very useful representation of the invariant measure 
$\mu_\epsilon$ in terms of embedded Markov chains, see Proposition \ref{kash} below. Then we 
prove the large deviations estimates. 
Let $U$ and $V$ be open subset of $X$ with compact closure with ${\overline U} \subset V$. 
Below, $U$ and $V$ will be the disjoint union of small neighborhoods of the sets $K_i$.
We introduce an increasing sequence of Markov times $\tau_0, \sigma_0, \tau_1, \dots $
defined as follows. We set $\tau_0=0$ and 
\begin{eqnarray}
\sigma_n \,&=&\, \inf \{ t > \tau_{n}\,:\, x_\varepsilon(t) \in \partial V
\} \label{mart1} \\
\tau_n \,&=&\, \inf \{ t > \sigma_{n-1}\,:\, x_\varepsilon(t) \in \partial U
\} \label{mart2} 
\end{eqnarray}
As a consequence of  hypoellipticity and the strong complete
controllability of the control problem associated to the diffusion
$x_\varepsilon(t)$ (condition {\bf L2} and {\bf L3})
we have the following result proven in \cite{Kl} which
extends to diffusions with hypoelliptic generators 
the characterization of invariant
measures in terms of recurrence properties of the process
$x_\varepsilon(t)$ and which is standard for diffusions with elliptic generators. 

We recall that the diffusion $x_\varepsilon(t)$ is positive recurrent if 
\begin{enumerate} 
\item It is recurrent, i.e., for all $x \in X$ for all open set
$U\subset X$ one has
\begin{equation}
P^\varepsilon_x ( \calR_U )  \,=\, 1
\end{equation}
where $ \calR_U$ is the event given by  
\begin{equation}
 \calR_U = \{ x_\varepsilon(t_n)
\in U {\rm ~for~an~increasing~sequence~} t_n \rightarrow \infty\}\,.
\end{equation}
\item For all $x\in X$ and for all open sets $U\subset X$, one has
\begin{equation}
E^\varepsilon_x( \sigma_U) \,< \, \infty
\end{equation}    
where $\sigma_U  = \inf \{ t, x_\varepsilon(t) \in U \}$. 
\end{enumerate}
It is proven in \cite{Kl}, Theorem 4.1, that if the diffusion $x_\varepsilon(t)$ is 
hypoelliptic  and strongly completely controllable then  the diffusion admits a (unique)
invariant measure $\mu_\varepsilon$ if and only if $x_\varepsilon(t)$ is positive
recurrent. 
Clearly it follows from this result that, almost surely, the Markov
times $\tau_j$ and $\sigma_j$  defined in Eqs.~\eqref{mart1} and
\eqref{mart2} are finite. 

An important ingredient in the proof of the results in \cite{Kl} is the following 
representation of the invariant measure $\mu_\varepsilon$: 
Suppose $x_\varepsilon(0)=x \in \partial U$. Then $\{x_\varepsilon(\tau_j)\}$
is a Markov chain with a (compact) state space given by $\partial U $
and which admits an invariant measure  $l_\varepsilon(dx)$. The following
result relates the measure $l_\varepsilon$ to the invariant measure
$\mu_\varepsilon$, see e.g. \cite{Ha}, Chap. IV, Lemma 4.2. for a proof. 

\begin{proposition} \label{kash}
Let the measure $\nu_\varepsilon$ be defined as
\begin{equation}\label{nu}
\nu_\varepsilon(D) \,=\, \int_{\partial U } l_\varepsilon(dx) E^\varepsilon_x
\int_0^{\tau_1}  
{\bf 1}_D (x_\varepsilon(t)) dt \,,
\end{equation}
where $D$ is a Borel set and ${\bf 1}_D$ is the characteristic
function of the set $D$. Then one has
\begin{equation}
\mu_\varepsilon(D) \,=\, \frac{ \nu_\varepsilon(D) } {\nu_\varepsilon(X)} \,.
\end{equation}
\end{proposition}
Up to the normalization, the invariant measure $\mu_\varepsilon$ assigns
to a set $D$ a measure equals to the time spent by the process in $D$
between two consecutive hits on $\partial U$.

The proof of Theorem \ref{single} is quite long and will be split into a
sequence of Lemmas. The proof is based on the following ideas: As
$\varepsilon \rightarrow 0$ the invariant measure is more and more
concentrated on a small neighborhood of the critical set
$\cup_i K_i$. To estimate the measure of a set $D$ one uses the 
representation of the invariant measure
given in Proposition \ref{kash} where the sets $U$ and $V$ are 
neighborhoods of the sets $\{K_i\}$. 
Let $\rho >0$ and denote $B(i,\rho)$ the $\rho$-neighborhood of $K_i$ 
 and $B(\rho)=\cup_i B(i,\rho)$. Let $D$ be a regular open set such that 
${\rm dist}(\cup_iK_i, D) > 0$. We choose $\rho'$ so small that 
${\rm dist} (B(i,\rho'), B(j,\rho')) > 0$, for $i\not=j$ and 
${\rm dist} (B(i,\rho'), D ) > 0$, for $i=1,\dots, L$, and we 
choose $\rho >0$ such that 
$0 <  \rho < \rho'$. We set $U=B(\rho)$ and $V=B(\rho')$. 
We let $\sigma_0$ and $\tau_1$ be the Markov times defined in Eqs.~\eqref{mart1} 
and \eqref{mart2} and let $\tau_D$ be the Markov time  defined as follows: 
\begin{equation}
\tau_D\,=\, \inf \{ t\,:\, x_\varepsilon(t) \in  D \} \,.
\end{equation}

The first two Lemmas will yield an upper bound on $\nu_\varepsilon(D)$, 
the unnormalized measure given by Eq.~\eqref{nu}. The first Lemma shows that, for 
$\varepsilon$ sufficiently small, the probability that the diffusion wanders around 
without hitting $B(\rho)$ or $D$ is negligible. 

\begin{lemma} \label{upper1}
For any compact set $K$ one has
\begin{equation}
\lim_{T \rightarrow \infty} \limsup_{\varepsilon \rightarrow 0} \varepsilon
\log \sup_{x\in K} P^\varepsilon_x(  \min\{\tau_D \wedge \tau_1\}  > T) \,=\, -\infty
\end{equation}
\end{lemma}

\noindent
{\em Proof:} 
If $x \in D\cup B(\rho)$, $\tau_D \wedge \tau_1 =0$ and there is nothing to
prove. Otherwise consider the closed sets
\begin{equation}
F_T\,=\, \left\{ \phi \in \calC([0,T])\,:\, \phi(s) \notin D \cup
B(\rho), {\rm~for~all~} s\in [0,T] \right\}\,. 
\end{equation}
Clearly the event $\{\tau > T\}$ is contained in $\{ x_\varepsilon \in
F_T\}$. By Theorem \ref{LDP}, we have for all $T < \infty$, 
\begin{equation}
\limsup_{\varepsilon \to 0} \varepsilon \log \sup_{x\in K}
P^\varepsilon_x(x_\varepsilon \in 
F_T) \, \le \, - \inf_{x\in K} \inf_{\phi \in F_T} I_{x,T}
(\phi) \,.
\end{equation}  
In order to complete the proof of the Lemma it is enough to show that 
\begin{equation}\label{limit}
\lim_{T\to \infty} \inf_{x\in K} \inf_{\phi \in F_T} I_{x,T}(\phi) \,=\,
\infty \,.
\end{equation}
Let $\phi^x$ be the trajectory of \eqref{eqmo01} starting at $x \in K$. 
By condition {\bf L1}, $\phi^x$ hits $B(\rho/3)$ in a finite
time $t^x$. By the continuous dependence of $\phi^x$  on its initial
condition, there 
is an open set $W^x$ such that for all $y\in W^x$, $\phi^y$ hits
the set $B(2\rho/3)$ before $t^x$. Since $K$ is compact, there is $T$
such that, for all $x \in K$, $\phi^x$ hits $B(2\rho/3)$ before
$T$. Assume now that the identity \eqref{limit} does not hold. Then, for
some $M < \infty$, and every integer $n$, there is $\psi_n \in F_{nT}$
such that  $I_{nT}(\psi_n) \leq M$. Consequently, for some $\psi_{n,k}
\in F_T$, we have 
\begin{equation}
M\,\ge\, I_{nT}(\psi_n) \,=\, \sum_{k=1}^n I_T(\psi_{n,k}) \,\ge \, n
\min_{k=1}^n I_T(\psi_{n,k}) \,.
\end{equation}
Therefore there is a sequence $\phi_n \in F_T$ such that
$\lim_{n\to\infty} I_T(\phi_n) =0$. Since the set $\{ \phi\,:\,
I_{x,T}(\phi) \le 1, \phi(0)\in K \}$ is compact, $\phi^n$ has a
limit point $\phi \in F_T$. Since $I_T$ is lower semicontinuous,
we have $I_T(\phi)=0$ and therefore $\phi$ is trajectory of
\eqref{eqmo01}. Since $\phi \in F_T$, $\phi$ remains outside of
$B(2\rho/3)$ and this is a contradiction with the definition of $T$. 
This concludes the proof of Lemma \ref{upper1}. \qed

Instead of the quantities $V(K_i,K_j)$ and $V(K_i,z)$ defined in 
Eqs.~\eqref{vij}  and \eqref{viz}, it is useful to introduce the following 
quantities: 
\begin{eqnarray} 
{\tilde V}(K_i,K_j) &=& \inf_{T>0} \inf \left\{ I_{x,T}(\phi),\, \phi(0)\in K_i,  
\phi(T) \in K_j, \phi(t) \not\in \cup_{l\not=i,j} K_l \right\} \,,
\nonumber \\ 
{\tilde V}(K_i,z) &=&  \inf_{T>0} \inf \left\{ I_{x,T}(\phi),\, \phi(0)\in K_i,  
\phi(T)=x,  
\phi(t) \not\in \cup_{l\not=i} K_l \right\}. \label{viztilde}
\end{eqnarray}

The following Lemma will yield an upper bound on the on $\nu_\varepsilon(D)$, 
where $\nu_\varepsilon$ is the (unnormalized) measure given by Eq.~\eqref{nu}. 

\begin{lemma} \label{upper2}
Given $h>0$, for $0 < \rho < \rho'$ sufficiently small one has
\begin{eqnarray}
(i) && \limsup_{\varepsilon\to 0} \varepsilon \log \sup_{y\in \partial B(i,\rho')} 
P^\varepsilon_y (\tau_D < \tau_1 ) \,\le\, - (\inf_{z \in D} {\tilde V}(K_i,z)-h)\,, 
\\
(ii) && \limsup_{\varepsilon\to 0} \varepsilon \log \sup_{y\in \partial B(i,\rho')} 
P^\varepsilon_y (x_\varepsilon(\tau_1) \in  \partial B(j,\rho) ) 
\,\le\, - ({\tilde V} (K_i,K_j)-h) \,. \nonumber\\
\end{eqnarray}  
\end{lemma}

\noindent
{\em Proof:} We first prove item (i). If 
$\inf_{z \in D} {\tilde V}(K_i,z)= +\infty$ there is no curve connecting 
$K_i$ to $z\in D$ without touching the other $K_j$, $j\not=i$.. Therefore 
$P^\varepsilon_y (\tau_D < \tau_1 )=0$ and there is nothing to prove.
Otherwise, 
for $h > 0$ we set ${\tilde V}_h  = \inf_{z\in D}
{\tilde V}(K_i,z) - h $. Since $V(y,z)$ satisfies the triangle inequality, 
we have, by condition {\bf L2}, that, for
$\rho$ small enough  
\begin{equation}
\inf_{y\in \partial B (i,\rho')} \inf_{z\in D} {\tilde V}(y,z) \, \ge \, 
\inf_{y\in \partial B (i,\rho')} \inf_{z\in D} {\tilde V}(K_i,z) 
- \sup_{y \in \partial B (i,\rho')} {\tilde V} (K_i,y) \, \ge\, {\tilde V}_h \,.
\end{equation} 
where 
\begin{equation}
{\tilde V}(y,z) \,=\, \inf_{T>0} \inf \left\{ I_{x,T}(\phi),\, \phi(0)=y,  \phi(T)=z, 
\phi(t) \not\in \cup_{l\not=i} K_l \right\} \,.
\end{equation}
By Lemma \ref{upper1}, there is $T < \infty$ such that 
\begin{equation} \label{f1}
\limsup_{\varepsilon \to 0} \varepsilon \log \sup_{y \in \partial B(i,\rho')} 
P^\varepsilon_y( \tau_D \wedge \tau_1 > T)  \,<\, - V_h \,. 
\end{equation}
Let $G_T$ denote the subset of $\calC([0,T])$ which consists of functions 
$\phi(t)$ such that $\phi(t) \in {\overline D}$ for some $t \in [0,T]$ and 
$\phi(t) \not\in  B(\rho)$ if $t \le \inf\{s \,, \phi(s) \notin D\}$. 
The set $G_T$ is closed as is  seen by considering its complement. 

We have 
\begin{equation}
\inf_{y\in \partial B(i,\rho')} \inf_{\phi \in G_T} I_{y,T}(\phi)\,\ge \,  
\inf_{y\in \partial B(i,\rho')} \inf_{z \in {\overline D}} {\tilde V}(y,z)\,\ge\,V_h\,, 
\end{equation}
and thus by Theorem \ref{LDP}, we have
\begin{equation} \label{f2}
\limsup_{\varepsilon \to 0} \varepsilon \log \sup_{y \in \partial B(i,\rho') }
P^\varepsilon_y(x_\varepsilon \in G_T) \,\le \, - \inf_{ y \in \partial B(i,\rho')}
\inf_{\phi \in G_T} I_{y,T}(\phi) \,\le\, - V_h  \,.
\end{equation}
We have the inequality
\begin{equation}
P^\varepsilon_y( \tau_D < \tau_1 ) \le P^\varepsilon_y (\tau_D \wedge \tau_1 > T) +
P^\varepsilon_y (x_\varepsilon \in G_T)\,,
\end{equation} 
and combining the estimates \eqref{f1} and \eqref{f2} yields
\begin{equation}
\limsup_{\varepsilon \to 0} \varepsilon \log \sup_{y \in \partial B(i,\rho')} 
P^\varepsilon_y( \tau_D \wedge \tau_1 ) \, \le \, - V_h \,. 
\end{equation}
This completes the proof of item (i) of Lemma \ref{upper2}. 

The proof of part (ii) of the Lemma is very similar to the first part and 
follows closely the corresponding estimates in \cite{FW}, Chapter 6, 
Lemma 2.1. The details are left to the reader. \qed

The following Lemma will yield  a lower bound on $\nu_\varepsilon(D)$. It
makes full use of the information contained in Lemmas \ref{continuity}
and \ref{cont}.

\begin{lemma} \label{lower1} 
Given $h > 0$, for $0 < \rho' < \rho$ sufficiently small  one has
\begin{eqnarray}
(i) && \liminf_{\varepsilon \to 0} \varepsilon \log \inf_{x\in \partial B (i,\rho)} 
P^\varepsilon_x( \tau_D <  \tau_1) \, \ge \, 
- ( \inf_{z\in D} {\tilde V}(K_i, z) + h) \,. \\
(ii) && \liminf_{\varepsilon \to 0} \varepsilon \log \inf_{x\in \partial B (i,\rho)} 
P^\varepsilon_x( x_\varepsilon(\tau_1) \in \partial B (j,\rho) ) \, \ge \, 
- ( {\tilde V}(K_i, K_j) + h) \,. \nonumber \\
\end{eqnarray}
\end{lemma}

\noindent
{\em Proof:} We start with the proof of item (i). If 
$\inf_{z\in D} {\tilde V}(K_i, z) = + \infty$ there is nothing to prove. 
Otherwise let $h>0$ be given. By condition {\bf L2}, (see  Corollary \ref{cont}), 
there are $\rho$ and
$\rho'>0$ with $\rho<\rho'/3$ and $T_0 < \infty$ such that, 
for all $x\in \partial B(i,\rho)$, there is a path $\psi^x \in \calC([0,T_0])$ 
which satisfies $I_{x,T_0}(\psi^x)\le h/3$ with $\psi^x(0)=x$ and
$\psi^x(T_0)=x_0 \in K_i$ and $\psi^x(t) \in B(2\rho'/3)$, $0\le t \le
T_0$. 

By condition {\bf L3}, 
there are $z\in D$, $T_1 < \infty$ and $\phi_1 \in \calC([0,T_1])$ such
that $I_{x_0, T_1}(\phi_1)\le \inf_{z \in D} {\tilde V}(K_i,z) + h/3$
and $\phi_1(0)=x_0 \in K_i$ and $\phi_1(T_1)= z $ and $\phi_1$ does not touch 
$K_j$, with $j\not=i$. We may and will assume that $\rho$ and $\rho'$ are chosen 
such that $2\rho' \le {\rm dist}(\phi_1(t), \cup_{j\not=i} K_j$.  
We note $\Delta={\rm dist} (z, \partial D)$. 
Let $x_1$ be the point of last intersection of
$\phi_1$ with $\partial B(i,\rho)$ and let $t_1$ such that
$\phi_1(t_1)=x_1$. We note $\phi_2 \in \calC([0,T_2])$, with
$T_2=T_1-t_1$, the path obtained from $\phi_1$ by deleting up to time
$t_1$ and translating in time. Notice that the path $\phi_2$ may hit
several times $\partial B(i,\rho')$, but hits $\partial B(i,\rho)$ 
only one time (at time $0$). 
Denote as  
\begin{equation}
\sigma=\inf\{t: \phi_2(t) \in \partial B (i,\rho')\} \label{sig}
\end{equation} 
the first time $\phi_2(t)$ hits $\partial B (i,\rho')$. We choose $\Delta'$ 
so small that if $\psi \in \calC([0,T_2])$ belongs to the $\Delta'$-neighborhood 
of $\phi_2$, then $\psi(t)$ does not intersect $\partial B (i,\rho)\}$ 
and $\partial B (i,\rho')\}$ for $0<t<\sigma$ and and does not intersect 
$\partial B (i,\rho)\}$ for $t> \sigma$.

%we note $\Delta'$ the 
%(positive) smallest distance between
%$\partial B(i,\rho)$ and $\phi_2(t)$ after the time $\sigma$
%\begin{equation}
%\Delta' \,=\, \inf_{\sigma \le t \le T_2} {\rm dist}
%(\phi_2(t),\partial B(i, \rho) \,>\,0 \,.  
%\end{equation} 

By condition {\bf L2}, there are $T_3< \infty$ and
$\phi_3 \in \calC([0,T_3])$ such that $\phi_3(0)=x_0$,
$\phi_3(T_3)=x_1$, $\phi_3(t) \in B(2\rho'/3)$, $0\le t \le
T_3$,  and $I_{x_0,T_3}(\phi_3) \le h/3$. 
Concatenating $\psi^x$, $\phi_3$ and $\phi_2$, we obtain a path $\phi^x \in
\calC([0,T])$ with $T=T_0+T_3+T_2$ and $I_{x,T}(\phi^x)\le
\inf_{z \in D} {\tilde V}(K_i,z) + h$. 
By construction the path $\phi^x$ avoids $\partial B (i,\rho)\}$
after the time 
$T_0+T_3+\sigma$ where $\sigma$ defined in Eq.~\eqref{sig}. 

We consider the open set
\begin{equation}
U_T\,=\, \bigcup_{x\in \partial B(\rho)} \left\{ \psi \in \calC ([0,T])\,:\,
\|\psi-\phi_x \| \le \min \{ \frac{\rho}{3}, \frac{\Delta}{2},
\frac{\Delta'}{2}\} 
\right\} \,.
\end{equation}
By construction the event $\{x_\varepsilon(t) \in U_T\}$ is contained in
the event $\{\tau_D \le \tau_1\}$. 
By Theorem \ref{LDP} we have 
\begin{eqnarray}
\liminf_{\varepsilon \to 0} \varepsilon \log \inf_{x\in \partial B(\rho)} 
P^\varepsilon_x( \tau_D < \tau_1) \, &\ge& \, 
\liminf_{\varepsilon \to 0} \varepsilon \log \inf_{x\in \partial B(\rho)} 
P^\varepsilon_x( x_\varepsilon \in U_T) \nonumber \\
\,&\ge& \, - \sup_{x \in \partial B(\rho) } \inf_{\psi \in U_T}
I_{x, T}(\psi) \nonumber \\ 
\,&\ge& \, - \sup_{x \in \partial B(\rho) } I_{x, T}(\phi^x)  \nonumber \\
\,&\ge& \, - (\inf_{z\in D} {\tilde V}(K_i, z) + h )\,.
\end{eqnarray} 
This concludes the proof of item (i). 

The proof of (ii) follows very closely the corresponding estimate in 
\cite{FW}, Chapter 6, Lemma 2.1., which considers the case where the 
generator of the diffusion is elliptic: for any $h>0$ one construct paths 
$\phi^{xy} \in \calC([0,T])$ from $x\in \partial B(i,\rho)$ to 
$y\in \partial B(j,\rho)$ such that 
$I_{x,T}(\phi^{xy}) \le {\tilde V}(K_i,K_j) + h/2$ and such that if 
$x_\varepsilon(t)$ is in a small neighborhood of $\phi^{xy}$, then 
$x_\varepsilon(\tau_1) \in \partial B(j,\rho)$. As in part (i) of the Lemma, 
the key element to construct the paths $\phi^{xy}$ is the condition 
{\bf L2} of small-time controllability around the sets $K_i$.  The details are left to 
the reader. 

This concludes the proof of lemma \ref{lower1}. \qed

The following two Lemmas give upper and lower bounds on the
normalization constant $\nu_\varepsilon(X)$, where $\nu_\varepsilon$ is
defined in Eq.~\eqref{nu}. 

\begin{lemma} \label{lower2}
For any $h > 0$, we have
\begin{equation}
\liminf_{\varepsilon \rightarrow 0} \varepsilon \log \nu_\varepsilon(X) \,\ge\, -h \,.
\end{equation}
\end{lemma}

\noindent
{\em Proof:} We choose an arbitrary $h>0$. For any $\rho' > 0$ we have
the inequality: 
\begin{eqnarray}
\nu_\varepsilon(X) \, &\ge&   \, \nu_\varepsilon(B(\rho')) \nonumber \\
                \, &= &   \, \int_{\partial B(\rho) } l_\varepsilon (dx)
E^\varepsilon_x \int_0^{\tau_1} {\bf 1}_{B(\rho')} (x_\varepsilon(t)) dt\nonumber \\ 
\, &\ge &   \, \int_{\partial B(\rho) } l_\varepsilon (dx)
E^\varepsilon_x \int_0^{\sigma_0} {\bf 1}_{B(\rho')} (x_\varepsilon(t)) dt
\nonumber \\
 \, &= &   \, \int_{\partial B(\rho) } l_\varepsilon (dx) E^\varepsilon_x(\sigma_0)
\nonumber \,.
\end{eqnarray}
We use a construction similar as that used in Lemma
\ref{lower1}. 
Using condition {\bf L2}, there are $\rho$ and $\rho'>0$ 
with $\rho<\rho'/3$ such that
such that for all $x \in \partial B(\rho) $, 
there are $T_1 < \infty$ and $\psi^x \in \calC([0,T_1])$ such that
$\psi^x(0)=x$, $\psi^x(T_1)=x_0\in \cup_i K_i$, 
$\psi^x(t)\in B(2\rho/3)$, $0\le t  \le T_1$
and $I_{x,T_1}(\psi^x) \le h/4$. 
Furthermore, using Corollary \ref{small}, for $\rho'$ small
enough, there are $z \in \partial B(\rho')$, $T_2<\infty$, and
$\psi \in \calC([0,T_2])$ such that $\psi(0)=x_0$, $\psi(T_2)=z$,
$\psi(t) \in  B(\rho')$  for $0\le t \le  T_2$,  
and $I_{x_0,T_2}(\psi) \le h/4$. We denote $\phi^x \in 
\calC([0,T])$, with $T=T_1+T_2$, the path obtained by concatenating
$\psi^x$ and 
$\psi$. It satisfies $I_{x,T}(\psi) \le h/2$. We consider the open set 
\begin{equation}
V_{T}\,=\, \bigcup_{x\in \partial B(\rho)} \left\{ \psi \in \calC ([0,T])\,:\,
\|\psi-\phi_x \| < \frac{\rho}{3} 
\right\} \,.
\end{equation}
Applying Theorem \ref{LDP}, one obtains
\begin{equation}
\liminf_{\varepsilon \to 0} \varepsilon \log \inf_{x\in \partial B(\rho)} 
P^\varepsilon_x(x_\varepsilon \in V_{T}) \, \ge \, - \frac{h}{2} \,.
\end{equation}
There is $T^*>0$, such that for any $\phi \in V_T$ the time
spent in $B(\rho')$ is at least $T^*$. Therefore, for $\varepsilon$ small
enough we obtain the bound
\begin{equation}
\int_{\partial B(\rho)} l_\varepsilon (dx) E^\varepsilon_x(\sigma_0) \, \ge T^*
\exp{(-\frac{h}{2\varepsilon})} \, \ge \, \exp{(-\frac{h}{\varepsilon})} \,, 
\end{equation}
and this completes the proof of Lemma \ref{lower2}.\qed

To get an upper bound on the normalization constant $\nu_\varepsilon(X)$
we will need an upper bound on the escape time out of the ball
$B(\rho')$ around $\cup_iK_i$, starting from $x \in \partial B(\rho)$.    

\begin{lemma} \label{upper3}
Given $h>0$, for $0 < \rho < \rho'$ sufficiently small, 
\begin{equation}
\limsup_{\varepsilon \to 0} \varepsilon \log \sup_{x\in \partial B(\rho)} 
E^\varepsilon_x(\sigma_0) \le h  \,.
\end{equation}
\end{lemma}

\noindent
{\em Proof:} Fix $h>0$ arbitrary. As in Lemma \ref{lower2}, we see
that, for $0 <\rho < \rho'$ sufficiently small and  for all $x\in
\partial B(\rho)$, 
there are $T_1,T_2 \le \infty$, $z \notin B(\rho') $ and $\phi^x \in
\calC([0,T_1+T_2])$  such that $\phi^x(0)=x$, $\phi^x(T_1)\in \cup_i K_i$,
and $\phi^x(T_1+T_2)=z$, and $I_{x,T_1+T_2}(\phi^x)\le h/2$. 
We set $T_0=T_1+T_2$ and $\Delta={\rm dist}(z,B(\rho'))$ 
and consider the open set 
\begin{equation}
W_{T_0}\,=\, \bigcup_{x\in \partial B (\rho)} \left\{ \psi \in \calC ([0,T_0])\,:\,
\|\psi-\phi_x \| < \frac{\Delta}{2}
\right\} \,,
\end{equation}
so that if $\psi \in W_{T_0}$ it escapes from $B(\rho')$ in a time less
than $T_0$, i.e., the event$ \{ x_\varepsilon \in W_{T_0} \}$ is contained in
the event $\{ \sigma_0 < T_0 \} $ . Using Theorem \ref{LDP}, we
see that for sufficiently small $\varepsilon$,
one has the bound
\begin{equation}
q\,\equiv\, \inf_{x\in \partial B(\rho)} P^\varepsilon_x( \sigma_0 < T_0 ) \,\ge\,
\exp{\left(-\frac{h}{2\varepsilon}\right) }\,. 
\end{equation}
Consider the events $\sigma_0 > kT_0$, $k=1,2,\dots$. Using the
Markov property one obtains the bound
\begin{eqnarray}
P^\varepsilon_x( \sigma_0 > (k+1)T_0 )\,&\le&\, \left[ 1- P^\varepsilon_x( kT_0 <
\sigma_0 \le (k+1)T_0)  \right] P^\varepsilon_x( \sigma_0 > kT_0) \nonumber\\ 
\,&\le&\, (1- q )  P^\varepsilon_x( \sigma_0 > kT_0). 
\end{eqnarray}
Iterating over $k$ yields
\begin{equation}
\sup_{x\in \partial B(\rho)} 
P^\varepsilon_x( \sigma_0 > kT_0 )\,\le\, (1- q )^k \,.
\end{equation}
Therefore
\begin{eqnarray}
\sup_{x\in \partial B(\rho)} E^\varepsilon_x( \sigma_0) \,&\le&\, T_0 \left[ 1 +
\sum_{k=1}^{\infty} \sup_{x\in \partial B(\rho) } 
P^\varepsilon_x( \sigma_0 > kT_0 )
\right] \nonumber \\ 
\,&\le&\, T_0 \sum_{k=0}^{\infty}(1-q)^k \,=\, \frac{T_0}{q}\,. 
\end{eqnarray}
Since $q \ge \exp{(-\frac{h}{2\varepsilon}) }$ one obtains, for sufficiently
small $\varepsilon $
\begin{equation}
\sup_{x\in \partial B(\rho)} E^\varepsilon_x( \sigma_0) \,\le\, T_0
\exp{\left(\frac{h}{2\varepsilon}\right)} \,\le \, 
\exp{\left(\frac{h}{\varepsilon}\right)} \,.
\end{equation}
This concludes the proof of Lemma \ref{upper3}. \qed

With this Lemma we have proved all large deviations estimates needed in the 
proof of Theorem \ref{single}.
We will need upper and lower estimates on 
$l_\varepsilon( \partial B(i,\rho))$ where $l_\varepsilon$ is the invariant measure 
of the Markov chain $x_\varepsilon(\tau_j)$. These estimates are proved in 
\cite{FW}, Chapter 6, Section 3 and 4 and are purely combinatorial and 
rely on the representation of the 
invariant measure of a Markov chain with a finite state space via graphs on 
the state space. 
By Lemma \ref{upper2}, (ii) and 
\ref{lower1}, (ii) we have the following estimates on the probability 
transition $q(x,y)$, $x,y \in \partial B(\rho)$ of the Markov chain 
$x_\varepsilon(\tau_j)$: Given $h>0$, for $0<\rho < \rho'$ sufficiently small 
\begin{equation}
\exp{-\frac{1}{\varepsilon}({\tilde V}(K_i,K_j) + h)}   
\,\le\, 
q(x, \partial B(j,\rho)) 
\,\le\,
\exp{-\frac{1}{\varepsilon}({\tilde V}(K_i,K_j) - h)}\,,   
\label{jk1}
\end{equation}
for all $x \in \partial B(i,\rho)$ and sufficiently small $\varepsilon$. It is 
shown in \cite{FW}, Chapter 6, Lemmas 3.1 and 3.2 that the bound \eqref{jk1} 
implies a bound on $l_\varepsilon( \partial B(i,\rho))$. One obtains
\begin{eqnarray}
&& \exp{\left(-\frac{1}{\varepsilon}({\tilde W}(K_i) - \min_j {\tilde W}(K_j) + h)\right) }
\,\le\, l_\varepsilon( \partial B(i,\rho)) \,\le\, \nonumber \\
&& \hspace{2cm}\,\le\,
\exp{ \left(-\frac{1}{\varepsilon} ({\tilde W}(K_i) - \min_j {\tilde W}(K_j) - h)\right)}
\label{li} 
\end{eqnarray}
for sufficiently small $\varepsilon$, where 
\begin{equation}
{\tilde W}(K_i) \,=\, 
\min_{g \in G\{i\}} \sum_{(m\rightarrow n) \in g} {\tilde V}(K_m,K_n) \,. 
\label{witilde}
\end{equation}
Also in \cite{FW}, Chapter 6, Lemma 4.1,  
${\tilde W}(K_i)$ is shown to be in fact equal to  $W(K_i)$ defined in 
Eq.~\eqref{wi}: 
\begin{eqnarray}
{\tilde W}(K_i) \,&=&\, 
\min_{g \in G\{i\}} \sum_{(m\rightarrow n) \in g} {V}(K_m,K_n)  \nonumber \\
\,&=&\, W(K_i)\,.
\end{eqnarray}
Furthermore it is shown Lemma 4.2 there  
that the function $W(x)$, defined by Eq.~\eqref{wx}, satisfies the identity
\begin{eqnarray}
W(x)\,&=&\, \min_{i} (W(K_i) + V(K_i,x)) - \min_j W(K_j) \nonumber \\ 
\,&=&\, \min_{i} ({\tilde W}(K_i) + {\tilde V}(K_i,x)) - \min_j {\tilde W}(K_j)
\label{wwtilde}
\end{eqnarray}

We can turn to the proof of Theorem \ref{single}. 

\noindent
{\em Proof of Theorem \ref{single}:}

In order to prove Eq.~\eqref{mm}, it is enough to show that, for any
$h>0$, there is $\varepsilon_0 >0$ such that, for $\varepsilon < \varepsilon_0$
we have the inequalities:
\begin{eqnarray}
\mu_\varepsilon(D) \,&\ge&\, \exp{ \left(- \frac{1}{\varepsilon}
(\inf_{z \in D} W(z) + h)\right) }
\label{lo} \,, \\
\mu_\varepsilon(D) \,&\le&\, \exp{ \left(- \frac{1}{\varepsilon}
(\inf_{z \in D} W(z) - h)\right)}
\label{up} \,.
\end{eqnarray}

We let $\rho' > 0$ such that 
$\rho' < {\rm dist}(x_{\min},D)$. 
Recall that $\tau_D = \inf \{ t\,:\, x_\varepsilon(t)\in D \}$ is the first hitting 
time of the set $D$. we have the following bound on the $\nu_\varepsilon(D)$
\begin{eqnarray} 
\nu_\varepsilon(D)
\,&\le&\, \sum_i l_\varepsilon (\partial B(i,\rho)) 
\sup_{x\in \partial B(i, \rho) }
E^\varepsilon_x \int_0^{\tau_1} {\bf 1}_D(x_\varepsilon(t)) dt \nonumber \\
\,&\le&\,
L \max_i l_\varepsilon (\partial B(i,\rho)) 
\sup_{x\in \partial B(i, \rho) } 
P^\varepsilon_x(\tau_D \le \tau_1)
\sup_{y\in \partial D} E^\varepsilon_y(\tau_1) \label{u1} \,.
\end{eqnarray}
By {\bf L1}, there exist a constant $C$ independent of 
$\varepsilon$ such that
\begin{equation} \label{u2}
\sup_{y\in \partial D} E^\varepsilon_y(\tau_1) \,\le \, C\,,
\end{equation}
for $\varepsilon \le \varepsilon_0$. 
From Lemma \ref{upper2}, (i),  given $h >0$, for sufficiently small $0<\rho<\rho'$, 
we have the bound
\begin{equation}\label{u3}
P^\varepsilon_x(\tau_D < \tau_1)\, \le \, \exp{ \left( -\frac{1}{\varepsilon}
(\inf_{z\in D} {\tilde V}(K_i,z) - h/4)\right)} \,,
\end{equation}
for sufficiently small $\varepsilon$.  
From Eq.~\eqref{li}, given $h>0$, for sufficiently small $0<\rho<\rho'$,
we have the bound 
\begin{equation}
l_\varepsilon( \partial B(i,\rho))
\,\le\,
\exp{\left( -\frac{1}{\varepsilon}({\tilde W}(K_i) - \min_j {\tilde W}(K_j) - h/4)\right) }
\label{u3'}
\end{equation}
From the estimates \eqref{u1}-\eqref{u3'}, and the identity \eqref{wwtilde}
we obtain the bound 
\begin{equation} 
\nu_\varepsilon(D) \, \le \,  \exp{ \left( -\frac{1}{\varepsilon} 
(\min_{z\in D} W(z)  + h/2) \right)} \,, \label{u27}
\end{equation}
for sufficiently small $\varepsilon$.  
From Lemma \ref{lower2}, given $h>0$, for sufficiently small $0<\rho<\rho'$, 
we have the bound
\begin{equation}\label{u4}
\nu_\varepsilon(X) \, \ge\, \exp{\left(-\frac{h}{2\varepsilon}\right)}\,,
\end{equation}
for sufficiently small $\varepsilon$.
Combining estimates \eqref{u27} and \eqref{u4}, we obtain that
\begin{equation} 
\mu_\varepsilon(D)\,\le\, \exp{\left( -\frac{1}{\varepsilon}(\inf_{z\in D} W(z) -h) 
\right)  }\,,
\end{equation} 
for sufficiently small $\varepsilon$ and this gives the bound \eqref{up}. 

In order to prove \eqref{lo}, we consider the set $D_{\delta}=\{x \in
D\,:\, {\rm dist}(x,\partial D) \ge  \delta \}$. For $\delta$
sufficiently small, 
$D_{\delta} \not= \emptyset$. By {\bf L3},
${\tilde V}(K_i,z)$ is upper semicontinuous in $z$ so that ${\tilde V}(K_i,z')
\le {\tilde V}(K_i,z) + h/4$, for $z'-z \le \delta$. Therefore 
\begin{equation}\label{l0}
\inf_{z\in D_\delta} {\tilde V}(K_i,z) \le \inf_{z\in D} {\tilde V}(K_i,z) + h/4\,.
\end{equation}
We have the bound
\begin{equation} \label{l1}
\nu_\varepsilon(D)\,\ge \, \max_i l_\varepsilon(\partial B(i,\rho)) 
\inf_{x \in \partial B(i,\rho)} P^\varepsilon_x (\tau_{D_\delta} < 
\tau_1 )  \inf_{x\in \partial D_\delta} E^\varepsilon_x \int_0^{\tau_1} {\bf
1}_D(x_\varepsilon(t)) dt \,.
\end{equation}
There is $\varepsilon_0> 0$ and a constant
${\overline C} > 0$   
such that we have the bound 
\begin{equation}\label{l2}
\inf_{x\in D_\delta} E^\varepsilon_x \int_0^{\tau_1} {\bf
1}_D(x_\varepsilon(t)) dt\, \ge \, {\overline C} \,>\, 0 \,,
\end{equation}
uniformly in $\varepsilon \le \varepsilon_0$. 
From Eq.~\eqref{li}, given $h>0$, for sufficiently small $0<\rho<\rho'$,
we have the bound 
\begin{equation}
l_\varepsilon( \partial B(i,\rho))
\,\ge\, \exp{\left( -\frac{1}{\varepsilon}({\tilde W}(K_i) - \min_j {\tilde W}(K_j) 
+ h/4) \right) }\,,
\label{l3}
\end{equation}
for sufficiently small $\varepsilon$. 
Furthermore, by Lemma \ref{lower1} and inequality \eqref{l0}, given $h
> 0$, for $0<\rho<\rho'$ sufficiently small, we have
\begin{equation}\label{l4}
\inf_{x\in \partial B(i,\rho)} P^\varepsilon_x (\tau_{D_\delta} \le \tau_1) \, \ge \,
\exp{\left(-\frac{1}{\varepsilon}(\inf_{z\in D} {\tilde V}(K_i,z)  + h/4) \right)}\,, 
\end{equation}
for sufficiently small $\varepsilon$. Combining estimates \eqref{l1}--\eqref{l4} 
and identity \eqref{wwtilde} we find
\begin{equation}\label{l23}
\nu_\varepsilon(D) \,\ge\, \exp{ \left( -\frac{1}{\varepsilon}(\inf_{z\in D} W(z) + h/2)
\right) }\,.
\end{equation}
In order to give an
upper bound on the normalization constant $\nu_\varepsilon(X)$, we use 
Eq.~\eqref{nu}. Using the Markov property, we obtain
\begin{eqnarray}
\nu_\varepsilon(X)\,&=&\, \int_{\partial B(\rho)} l_\varepsilon(dx) 
E^\varepsilon_x(\tau_1) \nonumber \\
\,&=&\,
\int_{\partial B (\rho)} l_\varepsilon(dx) 
\left( E^\varepsilon_x(\sigma_0) + E^\varepsilon_x 
( E^\varepsilon_{x_\varepsilon(\sigma_0)} (\tau_1))\right) \nonumber \\
\,&\le&\, \sup_{x\in \partial B(\rho)} E^\varepsilon_x(\sigma_0) + 
\sup_{y\in \partial B(\rho')} E^\varepsilon_y(\tau_1) \,. \label{l31}
\end{eqnarray}
By Lemma \ref{upper3}, given $h>0$, for sufficiently small $0<\rho<\rho'$ we have 
the estimate 
\begin{equation}
\sup_{x\in \partial B(\rho)} E^\varepsilon_x(\sigma_0) \, \le \,
\exp{\left(\frac{h}{2\varepsilon}\right)}\,,
\end{equation} 
for sufficiently small $\varepsilon$. 
By {\bf L1}, the second term on the right hand side of
\eqref{l31} is bounded by a constant, uniformly in $0\le \varepsilon \le
\varepsilon_0$. Therefore for we obtain the estimate
\begin{equation}\label{l66}
\nu_\varepsilon(X) \, \le \, \exp{\left( \frac{h}{2\varepsilon} \right)}\,,
\end{equation}
for sufficiently small $\varepsilon$.
Combining estimates
\eqref{l23}  and \eqref{l66} we obtain the bound
\begin{equation} 
\mu_\varepsilon(D)\,\ge\, \exp{\left( -\frac{1}{\varepsilon}( \inf_{z\in D}W(z) + h) 
\right) }\,,
\end{equation} 
and this is the bound \eqref{lo}. 
This concludes the proof of Theorem \ref{single}. \qed

%This proves the assertion of the
%Theorem in the case ${\rm dist}(U_K_i, D) >0$

%If ${\rm dist}(U_K_i, D) =0$, we consider the set 
%$D_\kappa=D \backslash B_{\kappa}$. 
%Using condition {\bf L2} and {\bf L3}, given $h>0$ 
%we can choose $\kappa$ so small that
% \inf_{z\in D} {\tilde V}(K_i,z) -h \,\le \,  
%\inf_{z\in D_\kappa} {\tilde V}(K_i,z) \,\le\, 
%\inf_{z\in D} {\tilde V}(K_i,z)

%Finally we consider the case $V_0=0$. In this case, ${\rm
%dist}(x_{min},D)=0$.  The upper bound
%\eqref{up} is trivial. For the lower bound, using Corollary
%\ref{small}, we can choose $\kappa$ so small that $\inf_{z\in
%D\backslash B_{\kappa}} V(x_min,z) \le h/2$. Because of the
%inequality
%\begin{equation}
%\mu_\varepsilon(D) \,\ge\, \mu_\varepsilon(D \backslash B_{\kappa})\,
%\end{equation}
%the lower bound \eqref{lo} for $V_0=0$ will follow from the case $V_0
%\not=0$.

%This concludes the proof of Theorem \ref{single}. \qed

\section{Properties of the rate function} \label{balance}

In this section we prove assertions \eqref{eqq} and \eqref{neqq} of Theorem 
\ref{maintheorem} and Proposition \ref{ddd}.
Recall that for Eq.~\eqref{eqmo005}, the rate function, 
$I^{(\eta)}_{x,T}(\phi)$, takes the following form: 
For $\phi(t)=(p(t),q(t),r(t))$, 
\begin{equation}
I^{(\eta)}_{x,T}(\phi) \,=\,  
\frac{1}{4\gamma \lambda^2} \int_0^T ({\dot r} +\gamma\lambda^2 \nabla_r G) D^{-1}
({\dot r} +  \gamma\lambda^2 \nabla_r G)
\label{rate}
\end{equation}
if  \begin{equation}
{\dot q} = \nabla_pG \,, \quad {\dot p} = -\nabla_q G\,, 
\end{equation}
and is $+\infty$ otherwise. 
Recall that for a path $\phi\in \calC([0,T])$ with $\phi(0)=x$ and $\phi(T)=y$
we denote ${\tilde \phi}$ the time reversed path which satisfy ${\tilde \phi}(0)=Jy$ and 
${\tilde \phi}(T)=Jx$.

\noindent
{\em Proof of Proposition \ref{ddd}:} We rewrite the rate function 
$I^{(\eta)}_{x,T}(\phi)$ as 
\begin{eqnarray}
&&I^{(\eta)}_{x,T}(\phi) =  \nonumber \\
&&\,=\,  \frac{1}{4\gamma\lambda^2} 
\int_0^T ({\dot r} +\gamma\lambda^2 \nabla_r G) D^{-1} 
({\dot r} +\gamma\lambda^2 \nabla_r G)dt \nonumber \\
&&=  \frac{1}{4\gamma\lambda^2} 
\int_0^T ({\dot r} - \gamma\lambda^2 \nabla_r G) D^{-1} 
({\dot r} -\gamma\lambda^2 \nabla_r G)dt
+ \int_0^T (\nabla_{r}G) D^{-1} {\dot r} dt  \nonumber \\
&&\equiv K_1(\phi) + K_2(\phi) \label{qw1}
\end{eqnarray}
The term $K_1(\phi)$ has the following interpretation:  
It is the rate function corresponding to the 
the set of stochastic differential equations
\begin{eqnarray}
dq \,&=&\, \nabla_p G  dt\,, \nonumber \\
dp \,&=&\, -\nabla_q G  dt\,, \nonumber \\
dr \,&=&\, + \gamma\lambda^2 \nabla_r G  dt  + 
\varepsilon^{1/2} (2\gamma \lambda^2 D)^{1/2} dw\,.  \label{eqmo50} 
\end{eqnarray}
In particular there is $u\in L^2([0,T])$ such that for 
$\phi(t)=(p(t),q(t),r(t))$, with $\phi(0)=x$ and $\phi(T)=y$ we have
\begin{eqnarray}
{\dot q}  \,&=&\, \nabla_p G  \,, \nonumber \\
{\dot p} \,&=&\, -\nabla_q G  \,, \nonumber \\
{\dot r}  \,&=&\, + \gamma\lambda^2 \nabla_r G   + 
 (2\gamma \lambda^2 D)^{1/2} u\,.  \label{eqmo52}
\end{eqnarray}
Consider now the transformation $(p,q,r) \rightarrow J(p,q,r)$ and 
$t \rightarrow -t$. This transformation maps the solution of Eq.~\eqref{eqmo52} 
into ${\tilde \phi}(t)=({\tilde p}(t),{\tilde q}(t),{\tilde r}(t))$ 
which is the solution to 
\begin{eqnarray}
{\dot {\tilde q}}  \,&=&\, 
\nabla_p G  \,, \nonumber \\
{\dot {\tilde p}} \,&=&\, 
-\nabla_q G  \,, \nonumber \\
{\dot {\tilde r}} \,&=&\, 
- \gamma\lambda^2 \nabla_r G  + (2\gamma \lambda^2 D)^{1/2} {\tilde u}(t)\,,  
\label{eqmo51} 
\end{eqnarray}
with  ${\tilde \phi}(0)= Jy$, 
${\tilde \phi}(T)=Jx$. 
This implies the equality
\begin{eqnarray}
K_1(\phi)\,&=&\, 
\frac{1}{4\gamma \lambda^2 }  \int_0^T ({\dot r} -\gamma\lambda^2 \nabla_r G)  
D^{-1} ({\dot r} -\gamma\lambda^2 \nabla_r G) \nonumber \\
\,&=&\, \frac{1}{4\lambda^2\gamma} \int_0^T ({\dot {\tilde r}} + \gamma\lambda^2 
\nabla_r G) D^{-1} ({\dot {\tilde r}} +\gamma\lambda^2 \nabla_r G) dt 
\nonumber \\
\,&=&\, I^{(\eta)}_{Jy, T} ({\tilde \phi})\, .
\end{eqnarray}
This means that $K_1(\phi)$ is nothing but the weight of the time reversed path, 
i.e. the path starting at time $0$ from $Jy$ and leading to $Jx$ at time $T$.
 
We now consider the second term $K_2(\phi)$ in Eq.~\eqref{qw1}. We consider 
separately the equilibrium case (i.e., $\eta=0$) and the non-equilibrium case 
(i.e. $\eta \not= 0$).  For $\eta =0$ the matrix $D$ is the identity and we find
\begin{eqnarray}
K_2(\phi)\,=\, \int_0^T \nabla_{r} G {\dot r} dt  \label{qw2}
\end{eqnarray}
Using the constraints ${\dot q}= \nabla_pG$ and 
${\dot p} =-\nabla_q G $ we obtain the identity $\nabla_{p} G {\dot p} + 
\nabla_{q} G {\dot q} = 0$ and therefore we get 
\begin{eqnarray}
\int_0^T \nabla_{r} G {\dot r} dt \,&=&\, 
\int_0^T \left(\nabla_{r} G {\dot r}  + \nabla_{p} G {\dot p} +  
\nabla_{q} G {\dot q}\right) dt \nonumber \\
\,&=&\, \int_0^T \frac{d}{dt} G dt\,=\, G(y)- G(x) \,,
\end{eqnarray}
and this proves Eq.~\eqref{dtbl} in the case $\eta=0$. 

To prove Eq.~\eqref{dtbl} in the case $\eta\not=0$ observe that we have the identity, 
\begin{equation}
\frac{d}{dt} \frac{1}{2}(\lambda^{-1}r_i - \lambda q_i)^2 \,=\, 
(\lambda^{-1}r_i - \lambda q_i)(\lambda^{-1}{\dot r}_i - \lambda{\dot q}_i)\,=\, 
\lambda \nabla_{r_i} G (\lambda^{-1}{\dot r}_i - \lambda p_i)\,,
\end{equation}
for $i=1,n$,  and therefore
\begin{equation}
\nabla_{r_i} G {\dot r}_i =  \lambda^{2} \nabla_{r_i} G p_i 
+ \frac{d}{dt} \frac{1}{2}(\lambda^{-1}r_i - \lambda q_i)^2 \,.
\end{equation}
Hence, using the definition \eqref{ep2}, we  obtain
\begin{equation}
K_2(\phi)\,=\, \int_0^T \nabla_{r} D^{-1} G {\dot r} dt  \,=\, 
R(\phi(T))- R(\phi(0)) - \int_0^T \Theta (\phi(t))  dt 
\end{equation}
This completes the proof of Proposition \ref{ddd}. \qed

With generalized detailed balance we show the following
\begin{proposition}\label{equilibriumaction} 
If $\eta=0$ then $W^{(0)}(x)=G(x) - \min_x G(x)$. 
\end{proposition}

\noindent
{\em Proof:} 
The function $W^{(0)}(x)$ is given by
\begin{equation}
W^{(0)}(x)\,=\, \min_{i} \left( W^{(0)}(K_i) + V^{(0)}(K_i,x) \right) - \min_j W^{(0}(K_j) \,.
\label{action}
\end{equation}
where the minimum is taken over all compact sets $K_i$.  In Eq.~\eqref{action}, 
$W^{(0)}(K_i)$ is given by
\begin{equation}
W^{(0)}(K_i)\,=\, \min_{g \in G\{i\}} \sum_{(m\rightarrow n) \in g} V^{(0)}K_m,K_n) \,. 
\label{transition}
\end{equation}  
The sets $K_j$ are the critical sets of the generalized Hamiltonian $G(p,q,r)$, 
therefore $G$ is constant on $K_j$ and we set $G(x)=G_j$ for all $x\in K_j$. 
Furthermore if  $(p,q,r) \in K_j$, then $p=0$ and therefore the sets $K_j$ are 
invariant under time reversal: $JK_j=K_j$. Using the generalized detailed balance, we 
see that for any path $\phi \in \calC([0,T])$ with $\phi(0)=x\in K_m$ and 
$\phi(t)=y \in K_n$ we have
\begin{equation}
I^{(0)}_{x,T}(\phi) \,=\, I^{(0)}_{Jy,T}({\tilde \phi}) + G(y)-G(x)  \,=\, 
I^{(0)}_{y,T}({\tilde \phi}) + G_n-G_m \,.
\end{equation}
Taking the infimum over all paths $\phi$ and all time $T$, we obtain the identity
\begin{equation}
V^{(0)}(K_m,K_n)\,=\, V^{(0)}(K_n,K_m) + G_m-G_n \,. 
\label{vmn}
\end{equation}
In Eq.~\eqref{transition} the minimum is taken over all $\{i\}$-graphs (see the 
paragraph above Theorem \ref{maintheorem} in the introduction).
Given an $\{i\}$-graph and a $j$ with $j\not=i$, 
there is a sequence of arrows leading from $j$ to $i$. Consider now the graph obtained 
by reversing all the arrows leading from $j$ to $i$; in this way we obtain a 
$\{j\}$-graph. Using the identity \eqref{vmn} the weight of this graph is equal to the weight 
of the original graph plus $G_j-G_i$. Taking the infimum over all graphs we obtain 
the identity
\begin{equation}
W^{(0)}(K_i)\,=\, W^{(0)}(K_j) + G_j -G_i\,,
\end{equation}
and therefore we have
\begin{equation}
W^{(0)}(K_i)\,=\, G_i + {\rm const}\,,
\end{equation}
and so
\begin{equation}
W^{(0)}(x)\,=\, \min_i (G_i + V^{(0)}(K_i,x)) -  \min_j G_j\,.
\label{action2}
\end{equation}
The second term in Eq.~\eqref{action2} is equal to $\min_x G(x)$, since $G(x)$ is 
bounded below. 

We now derive upper and lower bounds on the first term in 
Eq.~\eqref{action2}. A lower bound follows easily from Proposition \ref{ddd}: For any path 
$\phi \in \calC([0,T])$ with $\phi(0)=z \in K_i$ and $\phi(T)=x$ we obtain the inequality
\begin{equation}
I^{(0)}_{z,T}(\phi) \,=\, I^{(0)}_{Jx,T}({\tilde \phi}) + G(x)-G_i\, \ge \, G(x)-G_i \,,
\end{equation}
since the rate function is nonnegative. Taking infimum over all paths $\phi$ and time $T$
we obtain 
\begin{equation}
W^{(0)}(x)\,\ge \, G(x) - \min_x G(x)\,.
\end{equation}
To prove the lower bound we consider the trajectory 
${\tilde \phi}$ starting at $Jx$ at time $0$ which is the solution of the equation 
\begin{eqnarray}
{\dot q}_i &=&  \nabla_{p_i} G \qquad i=1,\cdots,n  \,,
\nonumber \\ 
{\dot p}_i &=& - \nabla_{q_i} G \qquad  i=1,\cdots,n
\,, \label{zeromm}  \\ 
{\dot r}_i &=& - \gamma \lambda^2 \nabla_{r_i} G
\qquad i=1,n \,. \nonumber  
\end{eqnarray}

By Lemma \ref{omegalemma}, there is some $K_j$ such that
$\lim_{t\rightarrow \infty} {\tilde \phi}(t) \in K_j$. 
Furthermore, since ${\tilde \phi}$ is a solution of Eq.~\eqref{zeromm}, the rate 
function of this path vanishes $I^{(0)}_{Jx,T}({\tilde \phi})=0$, for any $T>0$. 
Note that an infinite amount of time is needed to reach  $K_j$ in general. Now consider 
the time reversed path $\phi(t)$. It starts at $t=-T$ with $T\le \infty$ 
at $K_i$ and reaches $x$ at time $0$. For such a path we have 
\begin{equation}
\lim_{T\rightarrow \infty}I^{(0)}_{z,T}(\phi) \,=\, \lim_{T\rightarrow \infty} 
I^{(0)}_{Jx}({\tilde \phi}) + G(x) - G_i \,=\, G(x)-G_i\,,
\end{equation} 
and therefore 
\begin{equation}
V^{(0)}(K_i,x)\,\le\, G(x)-G_i\,.
\end{equation}
We finally obtain 
\begin{equation}
W^{(0)}(x) \, \le \, G_i + V^{(0)}(K_i,x) - \min_x G(x) \,\le\, G(x) - \min_x G(x) 
\end{equation}
and this concludes the proof of Proposition \ref{equilibriumaction}. \qed

We have the following bound on the rate function: 

\begin{lemma} If $\eta \ge 0$ then for any $\phi \in \calC([0,T])$, 
\begin{equation}
(1+\eta)^{-1}I^{(0)}_{x,T}(\phi) \, \le \, I^{(\eta)}_{x,T}(\phi)  \, \le \, 
(1-\eta)^{-1} I^{(0)}_{x,T}(\phi)\,,
\label{bound}
\end{equation}
and a similar bound for $\eta\le 0$. 
\end{lemma}

\noindent
{\em Proof:} 
The proof follows from the fact 
that the subset of $\calC([0,T])$ on which $I^{(\eta)}_{x,T}(\phi) < \infty$ is 
independent of $\eta$. This is seen from the definition of rate function \eqref{ratef}.
Inspection of Eq.~\eqref{rate} implies the bound  \eqref{bound}. \qed

From this we obtain immediately

\begin{corollary} If $\eta \ge 0$ then 
\begin{equation}
(1+\eta)^{-1}(G(x)-\min_x G(x)) \,\le\, W^{(\eta)}(x) \,\le\, 
(1-\eta)^{-1}(G(x)-\min_x G(x)) \,.
\end{equation}
and a similar bound for $\eta\le 0$. 
\end{corollary}

This concludes the proof of Theorem \ref{maintheorem}. 

\begin{acknowledgements} 
We would like to thank J.-P. Eckmann, M. Hairer, J. Lebowitz, 
C.-A. Pillet, and H. Spohn for useful discussions. 
This work was partially supported by Swiss National Science
Foundation (L.R.-B.) and NSF grant DMS 980139 (L.E.T).
\end{acknowledgements}

\end{document}